\documentclass[preprint,showpacs,aps]{revtex4}
\usepackage{graphicx}
\usepackage{amsmath,amssymb}
\begin{document}

\title{Theory of spatially inhomogneous Bloch oscillations in semiconductor superlattices}

\author{ L. L. Bonilla, M. \'Alvaro, M. Carretero}

\affiliation{Gregorio Mill\'an Institute for Fluid Dynamics, Nanoscience and Industrial Mathematics, Universidad Carlos III de Madrid, 
Avenida de la Universidad 30, 28911 Legan\'es, Spain}

\begin{abstract}
In a semiconductor superlattice with long scattering times, damping of Bloch oscillations due to scattering is so small that nonlinearities may compensate it and Bloch oscillations persist even in the hydrodynamic regime. To demonstrate this, a Boltzmann-Poisson transport model of miniband superlattices with inelastic collisions is proposed and hydrodynamic equations for electron density, electric field and the complex amplitude of the Bloch oscillations are derived by singular perturbation methods. For appropriate parameter ranges, numerical solutions of these equations show stable Bloch oscillations with spatially inhomogeneous field, charge, current density and energy density profiles. These Bloch oscillations disappear as scattering times become sufficiently short. For sufficiently low lattice temperatures, Bloch and Gunn type oscillations mediated by electric field, current and energy domains coexist for a range of voltages. For larger lattice temperatures (300 K), there are only Bloch oscillations with stationary amplitude and electric field profiles. 
\end{abstract}

\bigskip

\noindent {\it Date}: {\today}\\


\maketitle

\renewcommand{\thefootnote}{\arabic{footnote}}
\newcommand{\fin}{\newline \rule{2mm}{2mm}}

\setcounter{equation}{0}
\section{Introduction}
\label{sec:1}
Bloch oscillations (BOs) are coherent oscillations of the position of electrons inside energy bands of a crystal under an applied constant electric field $-F$. Their frequency is proportional to the field $F$ and to the lattice constant $l$: $\omega_{\rm Bloch}=eFl/\hbar$. BOs were predicted by Zener in 1934 as an immediate consequence of the Bloch theorem \cite{zener}, but they were not experimentally found until much later \cite{fel92}. For BOs to be observable in an experiment, their periods have to be shorter than the scattering time $\tau$, and therefore the applied field has to surpass $\hbar/(el\tau)$. This value is too large for most natural materials, in which $l$ is of angstrom size. In 1970, Esaki and Tsu suggested to create an artificial crystal, which they called a superlattice (SL), by growing many identical periods comprising a number of layers of two different semiconductors with similar lattice constants \cite{esaki}. The period of the resulting one dimensional crystal may be much larger, say about 10 nm, and this gives reasonable electric fields of about 10 kV/cm, which are within the range of experimental observation. Damped Bloch oscillations were first observed in 1992 in semiconductor SLs whose initial state was prepared optically \cite{fel92}. In recent years, BOs have been observed in other artificial crystals such as atoms placed in the potential minima of a laser-induced optical standing wave \cite{dah96}, photons in a periodic array of waveguides \cite{per99} and Bose-Einstein condensates in optical lattices \cite{gre01} among other systems \cite{leo}.

BOs are potentially important to design infrared detectors, emitters or lasers which can be tuned in the Terahertz frequency range simply by varying the applied electric field \cite{leo}. Another application is based on the fact that BOs give rise to a resonance peak in the absorption coefficient under dc+ac bias and a positive gain at THz frequencies \cite{ktitorov}. The latter has been observed in quantum cascade laser structures \cite{ter07}. These applications are severely limited due to scattering which rapidly damps BOs and, for a dc voltage biased SL, favors the formation of electric field domains (EFDs) whose dynamics yields self-sustained oscillations of lower frequency (GigaHertz) \cite{hof96,BGr05} (a phenomenon similar to the Gunn effect in bulk GaAs \cite{kroemer}). EFD formation may also preclude THz gain in simple dc+ac driven SL which is typically calculated assuming spatially uniform solutions of drift-diffusion or Boltzmann type equations \cite{leo,kro00a,kro00b,ale06,IRo76}. This assumption has not been tested by solving space-dependent equations with appropriate boundary conditions or by experiments in semiconductor superlattices. An interesting idea for efficient terahertz harmonics generation is to excite relaxation oscillations in the superlattice by incident radiation from a waveguide \cite{ign11}.

To understand the role of EFD formation in the observation of BOs or THz Bloch gain, our starting point should be a model in which BOs and EFDs are both possible solutions of the governing equations. One simple possibility is to consider single miniband SLs described by Boltzmann-Poisson transport equations. In 1971, Ktitorov, Simin and Sindalovskii (KSS) considered a one-dimensional Boltzmann transport equation with a collision model comprising two terms: a simple relaxation to equilibrium and another term describing energy conserving impurity collisions \cite{ktitorov}. Later Ignatov and Shashkin  replaced relaxation to a Boltzmann type local equilibrium proportional to the instantaneous electron density instead of relaxation to global equilibrium \cite{ISh87}. More general relaxations to Fermi-Dirac local equilibria and self-consistent coupling of electric field and electron density via the Poisson equation have been considered recently \cite{BEP03,elena,AB11}. The characteristic equations of these Boltzmann-Poisson models exhibit BOs as solutions and there is a hydrodynamic regime for large applied electric fields described by drift-diffusion equations \cite{BEP03}. However both the Boltzmann-Poisson and the drift-diffusion systems do not have BOs as solutions. Instead, they display self-sustained oscillations of the current through the SL due to periodic nucleation and motion of EFDs \cite{BEP03,elena}, similar to the Gunn effect in bulk GaAs \cite{kroemer}. The reason of this shortcoming becomes clear when  the equations for the moments of the distribution function are analyzed. It turns out that the current density and the energy density oscillate at the Bloch frequency during BOs but the electron density varies slowly. Thus a local equilibrium that depends only on the electron density (and not on the instantaneous value of the current and energy densities) cannot produce equations for these magnitudes in the hydrodynamic limit. The situation is similar to that found in gases which motivated the Bhatnagar-Gross-Krook (BGK) collision model \cite{bgk}. If we want to derive hydrodynamic equations for the mass density, average velocity and temperature of a gas using relaxation to local equilibrium as a collision term, the local equilibrium distribution must depend on these magnitudes \cite{bgk}. In this paper, we study a collision model similar to BGK's, i.e., relaxation to a local equilibrium depending on the electron, current and energy densities \cite{BC08}. {\em The most important property of the proposed model is that it allows the local equilibrium distribution to oscillate at the Bloch frequency,} which is the crucial feature (missing in the KSS kinetic equation) if we want to derive a hydrodynamic regime that allows BOs. Since the scattering processes in a SL dissipate energy and momentum, our collision model includes two nonzero restitution coefficients. This is similar to the case of low density granular flows in which inter-grain collisions preserve momentum but dissipate energy and the corresponding BGK model includes one restitution coefficient \cite{BMD}. In low density granular flows, inelastic collisions dissipate energy and, as a consequence, the granular gas is cooling down continuously (unless there is a continuous injection of energy, for example through the boundaries). Nevertheless using a Chapman-Enskog method, it is possible to derive hydrodynamic equations about a local equilibrium with a temperature that is continuously decreasing, the so-called homogeneous cooling state \cite{BMD}. The hydrodynamic equations derived in \cite{BMD} for the simple dissipative BGK model of low density granular flows have also been obtained for the inelastic Boltzmann equation (with an integral collision kernel) in a double limit of small Knudsen number and almost elastic collisions \cite{sel98}.

In this paper we derive and solve numerically hydrodynamic equations containing BOs and EFDs among their solutions for a dc voltage biased SL. Bloch gain for a dc+ac driven SL will be studied elsewhere. Hydrodynamic equations are derived in a double limit: (i) the field-dependent term and the collision term are of the same order and dominate all others in the kinetic equation, and (ii) the collisions are almost elastic so that energy and momentum dissipation are of the same order as the spatial gradients in the balance equations. Extensions of classical kinetic theory methods based on assumption (i), such as the Chapman-Enskog technique, yield transport coefficients which become singular if the electric field becomes zero. Fixing this shortcoming requires matching to a multiple time scales expansion based on assumption (ii) and on a quasi-stationary solution of the equations for the first moments of the distribution function. Our techniques might be useful in other problems in kinetic theory having a similar structure. 

Once these difficulties are overcome, we can show that, in the appropriate limit, the electron current density and mean energy oscillate at the Bloch frequency, whereas the electron density, the electric field and the envelope of the BOs vary on a slower scale and are described by balance equations (hydrodynamic regime). Appropriate boundary and initial conditions include initiation of the BOs possibly by optical means \cite{fel92}. Numerical solutions in the appropriate parameter range show that initial profiles for the field and the BO amplitude evolve to stable spatially inhomogeneous profiles at room temperature \cite{BAC11}. At low temperature (70 K), we have found that {\em Bloch oscillations and Gunn-type oscillations due to EFD dynamics may coexist}. Increasing lattice temperature produces large diffusion coefficients in the electron current density (averaged over the BOs) as compared to the convective part thereof. This eliminates the Gunn-type oscillations. At low lattice temperature, the diffusion does not change that much, but convection dominates the average electron current density, thereby facilitating movable EFDs and Gunn-type oscillations. This novel finding of coexisting BOs of about 0.36 THz and 13.8 GHz Gunn-type oscillations is somewhat unexpected as it is usually assumed that Gunn-type oscillations have to be eliminated to get BOs or THz gain \cite{kro00b}.

The rest of the paper is as follows. In Section \ref{sec:2}, we review the KSS-Poisson transport equations. These equations are written in nondimensional form in Section \ref{sec:3} and equations for the two first moments of the distribution function (electron, current and energy densities) are derived from them. The moment equations do not form a closed set because they depend on the second moment. Assuming that the second moment is a function of the lower moments to be found later (the closure expression), we derive a reduced system of modulation equations for electron density, electric field and amplitude of the BOs by means of a nonlinear multiple scales method. Its analysis shows that BOs are always damped for the KSS-Poisson model. In Section \ref{sec:4}, we present our dissipative BGK collision model whose local equilibrium distribution depends on electron, current and energy densities. The corresponding modulation equations may support self-sustained BOs as solutions. They contain closure functions of electron, current and energy densities that have to be calculated by a different method. The precise form of the closure expressions are found in Section \ref{app2} by matching the equations for electric field, electron density and amplitude of the BOs found in Section \ref{sec:3} to the result of using the Chapman-Enskog method (CEM) \cite{BT10} on a modified kinetic equation for a distribution function that depends on the BO phase. This method yields equations with transport coefficients which are singular in the limit of vanishing electric field but they are compatible with the modulation equations of section \ref{sec:3} in an intermediate limit of sufficiently small fields. This compatibility yields the sought closure expressions. The results of numerical simulations of the modulation equations with appropriate boundary and initial conditions are presented in section \ref{sec:6}. Section \ref{sec:7} contains our conclusions. \ref{app0} shows that Bloch oscillations are always damped for the dissipative BGK model with some particular local equilibrium distributions. \ref{app1} gives some technical details on the local Boltzmann equilibrium. In \ref{sec:5}, we derive a drift-diffusion system for the electric field in superlattices with strongly inelastic collisions by using a CEM similar to that described in Refs. \cite{BEP03,BT10}.

\section{The KSS Boltzmann-Poisson model}
\label{sec:2}
We shall present our ideas in the very simple case of a n-doped semiconductor 
SL having only one populated miniband with the tight-binding dispersion relation: 
\begin{eqnarray}
{\cal E}(k) = {\Delta\over 2}\, (1 - \cos kl),\quad v(k) = \frac{1}{\hbar}\, \frac{d{\cal E}}{dk}=\frac{\Delta l}{2\hbar}\, 
\sin kl.   \label{eq1}
\end{eqnarray}
Here $\Delta$ is the miniband width, $l$ the SL period, $\hbar$ is the Planck constant and $v(k)$ is the electron group velocity. Electron motion and the electric field are directed along the SL growth direction which we take as the $x$ axis. In this case, the following modified KSS model describes electron motion including impurity collisions (which conserve energy but not momentum) and inelastic electron-phonon collisions \cite{BEP03}:
\begin{eqnarray} 
&&\frac{\partial f}{ \partial t} + v(k)\, \frac{\partial f}{ \partial x}
+  \frac{eF}{ \hbar} \,\frac{\partial f}{\partial k} = Q_{e}(f) + Q_{p}(f),  
\label{eq2}\\
&&\varepsilon\, \frac{\partial F}{\partial x} = \frac{e}{ l}\, (n-N_{D}),  
\label{eq3}\\   
&& n = \frac{ l}{ 2\pi} \int_{-\pi/l}^{\pi/l} f(x,k,t)\, dk,    \label{eq4}\\
&& Q_{e}(f) = - \nu_{e}\,(f - f^{1D}),     \label{eq5}\\
&& f^{1D}(k;n) = \frac{m^{*}k_{B}T_{0}}{\pi\hbar^2}\, \ln\!\left[1+ \exp\!\left(
\frac{\mu - {\cal E}(k)}{k_{B}T_{0}}\right)\!\right] ,  \label{eq6}\\
&& \frac{ l}{ 2\pi} \int_{-\pi/l}^{\pi/l} f^{1D}(k;n)\, dk= n,    \label{eq7}\\
&& Q_{p}(f) = - \nu_p\,\mathcal{A}f\equiv -\frac{\nu_{p}}{2}\, [f(x,k,t) - f(x,-k,t)].    \label{eq8}
\end{eqnarray} 
Here $f$, $n$, $N_{D}$, $\varepsilon$, $k_{B}$, $-e<0$, $m^*$, $\mu$ and $-F=-\partial W/\partial x$ are the one-particle distribution function, the 2D electron density, the 2D doping density, the dielectric constant, the Boltzmann constant, the electron charge, the effective mass of the electron, the electro-chemical potential and the electric field, respectively. $W$ is the electric potential. Note that the 1D distribution functions have the same units as the 2D electron density $n$ and that the electrochemical potential $\mu$ is a function of $n$ obtained by solving (\ref{eq6})-(\ref{eq7}). The 1D Fermi-Dirac local equilibrium (\ref{eq6}) is obtained by integrating the 3D Fermi-Dirac distribution $1/(1+e^{[\mu-{\cal E}(\mathbf{k})]/(k_{B}T_{0})})$ with ${\cal E}(\mathbf{k})={\cal E}(k)+ \hbar^2\mathbf{k}_{\perp}^2/(2m^*)$ over the transversal wave vector $\mathbf{k}_{\perp}$. $T_{0}$ is the lattice temperature, $\nu_{e}$ and $\nu_{p}$ are collision frequencies which we take as given constants. The distribution function is periodic in $k$ with period $2\pi/l$. A quantum version of the semiclassical equations (\ref{eq2})-(\ref{eq8}) was studied in \cite{BE05}.

The KSS-Poisson system (\ref{eq2})-(\ref{eq8}) goes beyond relaxation to equilibrium and linear response theory. The collision terms in (\ref{eq2}) push the distribution function close to the local Fermi-Dirac equilibrium (\ref{eq6}) which depends on the instantaneous value of the electron density as indicated by (\ref{eq7}). In the case of a finite SL biased at zero volts and having insulating contacts, we can show that the system evolves toward a global equilibrium (\ref{eq6}) with $F=0$, $n=N_D$ and the chemical potential corresponding to this doping density by finding a free energy functional and using it to prove the H theorem \cite{AB11}. If the SL has Ohmic contacts and is subject to an appropriate dc voltage, Gunn type self-sustained oscillations are possible and the free energy oscillates at the same frequency \cite{AB11}.  

\subsection{Characteristic equations, moment equations and Bloch oscillations}
The characteristic equations associated to (\ref{eq2}) are
\begin{eqnarray}
&&\frac{dx}{dt}= v(k) = \frac{\Delta l}{2\hbar} \, \sin kl,
\quad \frac{dk}{dt} = \frac{eF}{\hbar},\label{eq9}
\end{eqnarray}
which, for constant $F$, have BO solutions $x(t)=-\frac{\Delta}{2eF}\cos\!\left(\frac{eFl}{\hbar}\, (t-t_0)\right)$. The influence of scattering can be seen from the equations for the moments of the distribution function. Since $\mathcal{E}(k)$ and $f^{1D}$ are even in $k$ and $v(k)$ and $\mathcal{A}f$ are odd in $k$, the collision operators $Q_e(f)$ and $Q_p(f)$ satisfy the conditions:
\begin{eqnarray}
&& \int_{-\pi/l}^{\pi/l} Q_{e,p}(f)\, dk = 0,  \quad
 \int_{-\pi/l}^{\pi/l} \mathcal{E}(k)\, Q_{p}(f)\, dk = 0,  \label{eq10}\\
&& \frac{e}{2\pi}\,\int_{-\pi/l}^{\pi/l} v(k)\, Q_{e,p}(f)\, dk = -\nu_{e,p} J_{n}, \label{eq11}\\
&&\int_{-\pi/l}^{\pi/l} \left[\frac{\Delta}{2}-{\cal E}(k)\right] Q_{e}(f)\, dk =-\nu_en(E-E^{1D}), \label{eq12}
\end{eqnarray} 
where 
\begin{eqnarray}
&&J_{n}(x,t)= \frac{e}{ 2\pi}\int_{-\pi/l}^{\pi/l} v(k)\, f(x,k,t)\, dk, \label{eq13}\\
&&E(x,t) = \frac{l}{ 2\pi\, n(x,t)} \int_{-\pi/l}^{\pi/l} \left[\frac{\Delta}{2} - {\cal 
E}(k)\right] f(x,k,t)\, dk,\label{eq14}\\
&&E^{1D}(x,t) = \frac{l}{ 2\pi\, n(x,t)} \int_{-\pi/l}^{\pi/l} \left[\frac{\Delta}{2} - {\cal 
E}(k)\right] f^{1D}(x,k,t)\, dk,\label{eq15}
\end{eqnarray}
are electronic current and energy densities. Thus $Q_e(f)$ dissipates energy and momentum whereas $Q_p(f)$ dissipates momentum but not energy. For a finite SL with insulating contacts and zero voltage bias, these collision terms dissipate the electron energy and momentum until the electrons reach equilibrium at the lattice temperature $T_0$, $n=N_D$, $F=0$ and zero current \cite{AB11}. 

To obtain equations for $n$, $J_n$ and $E$, we multiply (\ref{eq2}) by 1, $v(k)$ and $\Delta/2-{\cal E}(k)$, respectively, integrate over $k$ and simplify the results by means of (\ref{eq10})-(\ref{eq12}), thereby obtaining 
\begin{eqnarray} 
&& \frac{e}{l}\,\frac{\partial n}{\partial t} + \frac{\partial J_{n}}{\partial
x} = 0,  \label{eq16}\\
&& \frac{\partial J_{n}}{ \partial t} + \frac{e\Delta^2 l}{ 8\hbar^2}\,
\frac{\partial}{\partial x}(n-\mbox{Re}\, f_{2}) - \frac{e^2l\, nEF}{\hbar^2} = - 
(\nu_{e}+\nu_{p}) J_{n},    \label{eq17}\\
&& \frac{\partial E}{\partial t} - \frac{l E}{en}\,\frac{\partial J_{n}}{
\partial x} - \frac{\Delta^2l}{8\hbar n}\,\frac{\partial}{\partial x} \mbox{Im} 
f_{2} + \frac{F\, J_{n}l}{n} = - \nu_{e} (E-E^{1D}).   \label{eq18}
\end{eqnarray} 
Here we have used (\ref{eq1}) and the Fourier coefficients $f_{j}$ of the periodic distribution function:
\begin{equation}
f(x,k,t) = \sum_{j=-\infty}^\infty f_{j}(x,t)\, e^{ijkl}. \label{eq19}
\end{equation}
Note that $J_{n}= - e\Delta$ Im$f_{1}/(2\hbar)$ and $E= \Delta\, \mbox{Re} f_{1}/(2n)$. We can eliminate the electron density from (\ref{eq16}) by using the Poisson equation (\ref{eq3}) and integrating the result over $x$, thereby obtaining the following form of Amp\`ere's law
\begin{eqnarray} 
\varepsilon\,\frac{\partial F}{\partial t} + J_{n} = J(t).   \label{eq20}
\end{eqnarray} 
Here $J(t)$ is the total current density. Note that (\ref{eq16}) - (\ref{eq18}) are a closed system of equations in the case of space independent moments. The dissipation terms in the right hand side of (\ref{eq17}) and (\ref{eq18}) ensure that a global equilibrium $f=f^{1D}$ with $n=N_D$, $F=0$, $J_n=J=0$ and $E=E^{1D}$ is reached \cite{AB11}.

Note that space independent solutions of (\ref{eq16}) - (\ref{eq18}) with $\nu_e=\nu_p = 0$ (elastic collisions) have a constant electron density $n$, whereas $J_n$ and $E$ satisfy the equation of a linear oscillator with the Bloch frequency $\omega_{\rm Bloch}=eFl/\hbar$:
$$\frac{\partial J_n}{\partial t}-\frac{e^2lF}{\hbar^2}nE=0, \, \frac{\partial E}{\partial t}+\frac{lF}{n}J_n=0\Longrightarrow \frac{\partial^2 J_n}{\partial t^2}+\frac{e^2l^2F^2}{\hbar^2}J_n=0.$$

\section{Moment equations and damped BOs for the almost elastic KSS model}
\label{sec:3}
In this Section, we shall derive equations for the amplitude of the BOs in the limit of an almost elastic KSS transport equation. We shall use a quite general perturbation method that will be applied to other Boltzmann transport models later in this paper. Our results will show that BOs cannot be sustained within the KSS model and point out to the insufficiency thereof.

\subsection{Nondimensional KSS-Poisson and moment equations}
To study the KSS-Poisson transport equations and its associated moment equations, it is convenient to nondimensionalize them using the units indicated in Table \ref{t1}. They are appropriate for the hyperbolic limit $\delta\to 0$, in which the collision and Bloch frequencies are comparable and the corresponding terms dominate all others in (\ref{eq2}). Let $\nu$ be a typical collision frequency related to $\nu_e$ and $\nu_p$.  The field-dependent term in (\ref{eq2}) has the order $e[F]l[f]/\hbar$, whereas the collision terms are of order $\nu [f]$. Here $[f]$ and $[F]$ are typical scales of distribution function and field, respectively. Equations (\ref{eq3}) and (\ref{eq4}) with $[k]=1/l$ imply that $[f]=[n]=N_D$. Collision and field dependent terms are of the same order for $[F]=\hbar\nu/(el)$. From the Poisson equation (\ref{eq3}), we obtain: $[x]= \frac{\varepsilon [F] l}{eN_D}=\frac{\varepsilon\hbar\nu}{e^2N_D}$. The ratio from the convective term proportional to $[v(k)]=\Delta l/(2\hbar)$ to the collision term of order $\nu$ is a small dimensionless parameter
\begin{equation}
\delta=\frac{e^2N_D l\Delta}{2\varepsilon\hbar^2\nu^2}. \label{eq21}
\end{equation}
This is also the ratio between the scattering time and the dielectric relaxation time and it plays the same role as the Knudsen number in the kinetic theory of gases. Defining now $\hat{f}= f/N_D$, $\hat{n}=n/N_D$, $\hat{E}=2E/\Delta$, $\hat{J}_n=J/[J_n]$, $\hat{x}=x/[x]$, \ldots (where $[y]$ are the units in Table \ref{t1}), we can rewrite all equations so far written in nondimensional form. Omitting the hats over the variables, we find the following nondimensional versions of (\ref{eq2})-(\ref{eq8})
\begin{eqnarray} 
&&F\,\frac{\partial f}{\partial k}+ \frac{\nu_e}{\nu}\, (f-f^{1D})+\frac{\nu_p}{\nu}\,\mathcal{A} f=-\delta\left(\frac{\partial f}{ \partial t} + \sin k\, \frac{\partial f}{ \partial x}\right)\!,  \label{eq22}\\
&&\frac{\partial F}{\partial x} = n-1,  \label{eq23}\\   
&& n = \frac{1}{2\pi} \int_{-\pi}^{\pi} f(x,k,t)\, dk= \frac{1}{2\pi} \int_{-\pi}^{\pi} f^{1D}(x,k,t)\, dk,    \label{eq24}\\
&& f^{1D}(k;n) = \frac{m^{*}k_{B}T_{0}}{\pi\hbar^2N_D}\, \ln\!\left[1+ \exp\!\left(
\frac{2\mu -\Delta}{2k_BT_0} +\frac{\Delta}{2k_{B}T_{0}}\,\cos k\right)\!\right]\!.  \label{eq25}
\end{eqnarray} 

\begin{table}[ht]
\begin{center}\begin{tabular}{ccccccccc}
 \hline
$f$, $n$ & $F$ &${\cal E}$, $E$ &$v(k)$&$J_{n}$& $x$ & $k$ & $t$ & $\delta$\\
$N_{D}$ & $\frac{\hbar\nu}{el}$ & $\frac{\Delta}{2}$ & $\frac{l\Delta}{2
\hbar}$ & $\frac{eN_{D}\Delta}{2\hbar}$& $\frac{\varepsilon\hbar\nu}{
e^2N_{D}}$ & $\frac{1}{l}$ & $\frac{2\varepsilon\hbar^2\nu}{e^2N_{D}l
\Delta}$ & $\frac{e^2N_D l\Delta}{2\varepsilon\hbar^2\nu^2}$\\
$10^{10}$cm$^{-2}$ & kV/cm & meV & $10^4$m/s 
& $10^4$A/cm$^2$ & nm & 1/nm & ps & -- \\
$4.048$& 130 & 8 & 6.15 & 7.88 & 116 & 0.2 & 1.88& 0.0053\\
 \hline
\end{tabular}
\end{center}
\caption{Hyperbolic scaling and nondimensionalization with $\nu=10^{14}$ Hz.}
\label{t1}
\end{table}

The moment equations (\ref{eq16})-(\ref{eq18}) in nondimensional form are
\begin{eqnarray} 
&&\frac{\partial n}{\partial t} + \frac{\partial J_{n}}{\partial
x} = 0,  \label{eq26}\\
&& nEF=\delta\left[\frac{\partial J_{n}}{ \partial t} + \frac{1}{2}\,
\frac{\partial}{\partial x}(n-\mbox{Re}\, f_{2}) + \gamma_j J_{n}\right],    \label{eq27}\\
&&\frac{FJ_n}{n}=-\delta\left[ \frac{\partial E}{\partial t} - \frac{E}{n}\,\frac{\partial J_{n}}{\partial x} - \frac{1}{2 n}\,\frac{\partial}{\partial x} \mbox{Im} f_{2} + \gamma_{e} (E-E^{1D})\right],   \label{eq28}
\end{eqnarray} 
provided we define $\gamma_e$ and $\gamma_j$ through the relations
\begin{eqnarray} 
\frac{\nu_e}{\nu}=\delta\gamma_e, \quad \frac{\nu_e+\nu_p}{\nu}=\delta\gamma_j.
\label{eq29} 
\end{eqnarray}

We can rewrite (\ref{eq27}) and (\ref{eq28}) in terms of $f_1= nE-iJ_n$, the first harmonic of the distribution function: 
\begin{equation}
f(x,k,t;\delta)=\sum_{j=-\infty}^\infty f_j(x,t;\delta)\, e^{ijk}, \label{eq30}
\end{equation}
as
\begin{eqnarray} 
\left(\delta\,\frac{\partial }{\partial t}+ iF+ \delta\,\frac{\gamma_e+\gamma_j}{2}\right) f_1+ \delta\,\frac{\gamma_e-\gamma_j}{2}\overline{f_1}=\delta\gamma_enE_0 -\frac{\delta}{2i}\frac{\partial}{\partial x}(n- f_{2}),  \label{eq31}
\end{eqnarray} 
where $\overline{f_1}$ is the complex conjugate of $f_1$. The Amp\`ere law (\ref{eq20}) is simply
\begin{eqnarray} 
\frac{\partial F}{\partial t} + J_{n} = J.  \label{eq32}
\end{eqnarray} 

\subsection{Amplitude of the Bloch oscillations }
The moment equations (\ref{eq26}) and (\ref{eq31}) for $n=f_0$ and $f_1$ are not closed because the higher moment $f_2$ appears in them. In general, equations for moments $f_0$, \ldots, $f_n$ will contain terms depending on $f_{n+1}$. Singular perturbation methods, such as the CEM \cite{BT10}, produce a closure expression 
\begin{equation}
f_{2}= g(n,F,f_{1};\delta), \label{eq33}
\end{equation}
in which $g$ can be written as a power series in $\delta$. For the KSS-Poisson model, such an expression was derived in \cite{BEP03}. We will assume for the time being that $g$ is a given known function and derive modulation equations for the slowly varying quantities $n(x,t)$, $F(x,t)$ and $A(x,t)$. 

If we assume (as it is usually done in the method of multiple scales) that the moments and the field are functions of both a fast time scale $\tau=t/\delta$ (corresponding to a dimensional time unit $1/\nu$) and the slow time scale $t$, $n=n(x,\tau,t;\delta)$, $F=F(x,\tau,t;\delta)$ and $f_1=f_1(x,\tau,t;\delta)$, so that $\partial n/\partial t$ in (\ref{eq26}) becomes $\partial n/\partial t+\delta^{-1}\partial n/\partial\tau$ and so on. Equations (\ref{eq26}), (\ref{eq31})-(\ref{eq32}) should be replaced by
\begin{eqnarray} 
&& \frac{\partial n}{\partial \tau} =- \delta\left(\frac{\partial n}{\partial t}+\frac{\partial J_{n}}{\partial x}\right)\!,  \label{eq34}\\
&&\frac{\partial F}{\partial\tau}= \delta\left(J-J_{n}-\frac{\partial F}{\partial t}\right)\!, \label{eq35}\\
&&\left(\frac{\partial }{\partial\tau}+iF\right)\!f_1= \delta\!\left[\gamma_enE^{1D}-\frac{\gamma_e+\gamma_j}{2}f_1-\frac{\gamma_e-\gamma_j}{2}\overline{f_1} \right.\nonumber\\
&&\quad\quad\quad\quad\quad\quad -\frac{1}{2i}\frac{\partial}{\partial x}(n-g) \left.-\frac{\partial f_1}{\partial t}\right]\!.  \label{eq36}
\end{eqnarray} 
Setting now $\delta=0$, we find 
\begin{eqnarray} 
 n = n(x,t),\quad F=F(x,t), \quad f_1= A(x,t)\, e^{-iF\tau},  \label{eq37}
\end{eqnarray} 
in which $n(x,t)$, $F(x,t)$ and the envelope function $A(x,t)$ do not depend on the fast time scale. (\ref{eq34}) indicates that $n$ varies slowly on the time scale $t$. Similarly and according to (\ref{eq35}), $F$ is independent of $\tau$ provided the total current density $J(t)$ is of order 1. In practice, the size of $J$ is set by $J_{n}$ and by the bias condition. Imposing a voltage bias condition between contacts at the ends of a SL with finitely many periods, $J=O(1)$ if we assume that this voltage is constant or it varies on the slow scale $t$. We shall not consider in this paper the case of voltage bias varying on the fast time scale $\tau$ (e.g., an ac voltage biased SL driven at a frequency of order $1/\delta$), for which $J=O(1/\delta)$, and we have to modify the present analysis. 

The solution (\ref{eq37}) of (\ref{eq34}) - (\ref{eq36}) for $\delta=0$ exhibits BOs with frequency $F$. Before deriving modulation equations, it is useful to get first a {\em quasi-stationary distribution function} that solves (\ref{eq36}) and is independent of $\tau$:
\begin{eqnarray} 
f_{1,S} &=& \frac{\delta}{F^2+\delta^2\gamma_{j}\gamma_e}\left[\gamma_{e}
nE^{1D} (\delta \gamma_{j}-iF) + \frac{F+i\delta\gamma_{e}}{2}\, 
\frac{\partial}{\partial x}(n-\mbox{Re}\, g_{S}) \right.\nonumber\\
&+&\left. \frac{\delta\gamma_{j}-iF}{2}\,\frac{\partial}{\partial x} \mbox{Im}\, g_{S}+(iF-\delta\gamma_{j})\mbox{Re }h_{S}-(F+i\delta\gamma_{e}) \mbox{Im }h_{S}\right],  \label{eq38}
\end{eqnarray} 
provided we have replaced $h(x,t)=\partial f_{1}/\partial t$. We introduce the function $h(x,t)$ because extra terms having this form appear in the moment equations when we use the CEM. The specific expressions for $g_{S}$ and $h_{S}$ will be obtained by matching our results in this Section with those obtained by the CEM. See Section \ref{app2} and \ref{sec:5}. \bigskip

\noindent\textit{Remark 1.} All terms in Equation (\ref{eq38}) have nonzero limits as $F\to 0$ and this equation is the key step in the regularization of the results obtained in Section \ref{app2} using the CEM.
\bigskip

Eq.\ (\ref{eq38}) is equivalent to
\begin{eqnarray} 
 J_{n,S} &=& \frac{\delta}{F^2+\delta^2\gamma_{j}\gamma_e}\left[\gamma_{e}E^{1D}nF+\frac{F}{2}\,\frac{\partial}{\partial x}\mbox{Im}\, g_{S}\right.\nonumber\\
&-&\left. \frac{\delta\gamma_{e}}{2}\, \frac{\partial}{\partial x}(n-\mbox{Re}\, g_{S}) -F\,\mbox{Re}\, h_{S}+\delta\gamma_{e}\mbox{Im}\, h_{S}\right], \label{eq39}\\
E_{S}&=& \frac{\delta}{F^2+\delta^2\gamma_{j}\gamma_e}\left[\delta \gamma_{j}\left(\gamma_{e}E^{1D}+\frac{1}{2n}\, {\partial\over\partial x} \mbox{Im}\, g_{S}\right) \right. \nonumber\\
&&\quad\quad \quad\left. +\frac{F}{2n}\,\frac{\partial}{\partial x}
(n-\mbox{Re}\, g_{S})- \frac{\delta\gamma_{j}}{n}\,\mbox{Re}\, h_{S}- 
\frac{F}{n}\,\mbox{Im}\, h_{S}\right].   \label{eq40}
\end{eqnarray} 
The subscript S in $g_{S}$ and in $h_{S}$ stresses that these functions are calculated with $\tau$-independent $n$, $F$, $J_{n,S}$ and $E_{S}$. Note that for $F=O(1)$, $f_{1,S}=O(\delta/F)$, whereas $f_{1,S}=O(1)$ if $F\ll\delta$. Thus the order of $f_{1,S}$ depends on the order of magnitude of $F$ and it is better to treat the compact expression (\ref{eq38}) as an $O(\delta)$ quantity. Without the $x$-derivatives and $t$-derivative in the functions $g$ and $h$, the right hand sides of (\ref{eq39}) and (\ref{eq40}) correspond to the uniform stationary state 
\begin{eqnarray} 
&& J_{n,Su} = n\, v_d(F;n), \quad v_d=\frac{\delta\gamma_eE^{1D}F}{\delta^2 \gamma_j\gamma_{e}+F^2}, \quad E_{Su}= \frac{\delta^2\gamma_e\gamma_j E^{1D}}{\delta^2 \gamma_j\gamma_{e}+F^2}.  \label{eq41}
\end{eqnarray} 
The Boltzmann limit of (\ref{eq25}), corresponds to approximating $\ln(1+x)\approx x$ in that expression and then calculating the chemical potential by means of (\ref{eq24}). The resulting expression $f^{1D}\approx n\,\pi e^{\tilde{\beta}_0\cos k}/I_0(\tilde{\beta}_0)$, where $\tilde{\beta}_0=\Delta/(2k_BT_0)$ and $I_0(x)$ is the modified Bessel function of index zero \cite{AS}, produces a constant value $E^{1D}=I_1(\tilde{\beta}_{0})/I_0(\tilde{\beta}_0)$. Inserting this in the drift velocity, (\ref{eq41}) gives the well-known Ignatov-Shashkin formula \cite{ISh87} 
\begin{eqnarray} 
v_d(F)= \frac{2v_M \mathcal{F}}{1+ \mathcal{F}^2},\quad v_{M}=
\frac{\Delta l}{4\hbar\tau_e}\frac{I_{1}(\tilde{\beta}_{0})}{I_{0}(
\tilde{\beta}_{0})}, \quad \mathcal{F}=\frac{eFl}{\hbar\nu_e\tau_e},\quad  \tau_e=\sqrt{\frac{\nu_e+\nu_p}{\nu_e}},\label{eq42}
\end{eqnarray} 
which we have written back in dimensional units. It reduces to the Esaki-Tsu drift velocity in the limit $\tilde{\beta}_{0}\to\infty$ (zero lattice temperature), in which the Bessel functions are absent. \bigskip

\noindent \textit{Remark 2.} Comparing the Ignatov-Shashkin formula (\ref{eq42}) to experimentally obtained current--voltage characteristic curves yields the numerical values of the collision frequencies $\nu_e$ and $\nu_p$ \cite{sch98}. 
\bigskip

According to (\ref{eq40}), the mean energy $E$ decreases as the field $F$ increases, whereas the average energy $\langle \mathcal{E}\rangle$ obtained by averaging (\ref{eq1}),
\begin{eqnarray}
\langle \mathcal{E}\rangle=\frac{l}{2\pi n}\int_{-\pi/l}^{\pi/l}\mathcal{E}
\, f dk= \frac{\Delta}{2} - E, \label{eq43}
\end{eqnarray} 
increases with the electric field, as one would have expected. Note that (\ref{eq41}) is an asymptotically stable stationary solution of the moment equations (\ref{eq16})-(\ref{eq18}) provided we ignore the spatial dependence of $n$, $J_{n}$ and $E$. 

If we insert $f_{1}=f_{1,S}(x,t)+\Phi(x,t,\tau)$ in (\ref{eq36}), we obtain the equation:
\begin{eqnarray}
\left({\partial\over \partial\tau}+iF\right)\Phi=-\delta\,\left[\frac{
\gamma_{e}+\gamma_{j}}{2}\Phi+\frac{\gamma_{e}-\gamma_{j}}{2}\overline{\Phi} 
+\frac{\partial\Phi}{\partial t}-\frac{1}{2i} {\partial\over\partial x}(g-g_{S})\right]\!\!.  \quad\label{eq44}
\end{eqnarray} 
Since $F$ and $n$ are still varying on the slow time scale $t$, it is appropriate to introduce the following nonlinear fast time scale instead of $\tau$:
\begin{equation}
\theta =\frac{1}{\delta}\,\int_{0}^t F(x,s)\, ds, \label{eq45}
\end{equation}
which yields $\partial\theta/\partial t = F/\delta$, $\partial\theta/\partial\tau=F$. Note that, in dimensional units, the phase $\theta$ equals the integral of the Bloch frequency $eFl/\hbar$ over dimensional time, and therefore the partial derivative of $\theta$ over dimensional time equals the Bloch frequency. Thus $\theta$ is the phase of the Bloch oscillations.

The fast and slow time scales $\theta$ and $t$ will be used to set up a method of nonlinear multiple scales below in order to find out the modulation equations on the slow time scale $t$. If we consider $n$, $F$ and $\Phi$ to be functions of $x$, $\theta$ and $t$, Eqs.\ (\ref{eq34}), (\ref{eq35}) and (\ref{eq44}) become
\begin{eqnarray} 
 F{\partial n\over \partial\theta}&=&-\delta\,\left[\frac{\partial n}{\partial t}
-\frac{\partial}{\partial x}\mbox{Im }(f_{1,S} + \Phi)\right],\label{eq46}\\
F\left({\partial\over \partial\theta}+i\right)\Phi&=&-\delta\,\left[\frac{\partial\Phi}{\partial t} + \frac{\gamma_{e}+\gamma_{j}}{2}\Phi \right. \nonumber\\
&+& \frac{\gamma_{e}-\gamma_{j}}{2}\overline{\Phi} - \left. \frac{1}{2i} 
\frac{\partial}{\partial x}(g-g_{S})\right].  \label{eq47}
\end{eqnarray} 

The method of multiple scales is based on the expansions:
\begin{eqnarray}
n(x,t;\delta)= \sum_{m=0}^1\delta^m n^{(m)}(\theta,x,t)+ O(\delta^2),
\label{eq48}\\
F(x,t;\delta)= \sum_{m=0}^1\delta^m F^{(m)}(\theta,x,t)+ O(\delta^2),
\label{eq49}\\
\Phi(x,t;\delta)= \sum_{m=0}^1\delta^m \Phi^{(m)}(\theta,x,t)+ O(\delta^2),
\label{eq50}
\end{eqnarray}
and on assuming that $n^{(m)}$, $F^{(m)}$ and $\Phi^{(m)}$ are $2\pi$-periodic functions of $\theta$. Inserting (\ref{eq48}) - (\ref{eq50}) in (\ref{eq46}) - (\ref{eq47}) and (\ref{eq23}), we obtain the following hierarchy of equations: 
\begin{eqnarray} 
 \frac{\partial n^{(0)}}{ \partial\theta}&=&0, \label{eq51}\\
 \frac{\partial F^{(0)}}{\partial x}&=& n^{(0)}-1,\label{eq52}\\
F^{(0)}\left(\frac{\partial}{ \partial\theta}+i\right)\!\Phi^{(0)}&=& 0, \label{eq53}
\end{eqnarray}
\begin{eqnarray}
F^{(0)}\frac{\partial n^{(1)}}{ \partial\theta}&=&-\frac{\partial n^{(0)}}{\partial 
t}+\frac{\partial}{\partial x}\mbox{Im }(f_{1,S}+\Phi^{(0)}),\label{eq54}\\
 \frac{\partial F^{(1)}}{\partial x}&=& n^{(1)},\label{eq55}\\
F^{(0)}\left({\partial\over \partial\theta}+i\right)\!\Phi^{(1)}&=&-
\frac{\partial\Phi^{(0)}}{\partial t} - \frac{\gamma_{e}+\gamma_{j}}{2}
\Phi^{(0)} \nonumber\\
&+& \frac{\gamma_{e}-\gamma_{j}}{2}\overline{\Phi^{(0)}} + \frac{1}{2i} 
\frac{\partial}{\partial x}(g^{(0)}-g_{S}^{(0)}),\label{eq56}
\end{eqnarray} 
and so on.

The solution of (\ref{eq53}) is
\begin{equation}
\Phi^{(0)}= A(x,t)\, e^{-i\theta}, \label{eq57}
\end{equation}
whereas (\ref{eq51}) and (\ref{eq52}) indicate that $n^{(0)}$ and $F^{(0)}$ do not depend on $\theta$ \cite{footnote1}. The solutions of (\ref{eq54}) and (\ref{eq56}) are $2\pi$-periodic functions of $\theta$ only if the right hand sides of these equations do not contain secular terms proportional to 1 and $e^{- i\theta}$, respectively. This is the case if the integral of the right hand side of (\ref{eq54}) and the integral of $e^{i\theta}$ times the right hand side of (\ref{eq56}) over $[-\pi,\pi]$ are both zero. These solvability conditions give:
\begin{eqnarray}
&&\frac{\partial n^{(0)}}{\partial t} -\frac{\partial}{\partial x}\mbox{Im } f_{1,S} = 0,\label{eq58}\\
&&\frac{\partial A}{\partial t} = -\frac{\gamma_{e}+\gamma_{j}}{2}\, A
+\frac{1}{2i}\,\frac{\partial}{\partial x}\int_{-\pi}^\pi e^{i\theta}
g(n^{(0)},F^{(0)},f_{1,S}+A e^{-i\theta};0)\,\frac{d\theta}{2\pi}. \label{eq59}
\end{eqnarray}
Instead of (\ref{eq58}), we can use the Amp\`ere's law (\ref{eq32}) averaged over $\theta$ with $\langle J_n\rangle=-$ Im$f_{1,S}$ given by (\ref{eq39}):
\begin{eqnarray}
&&\frac{\partial F^{(0)}}{\partial t} + \frac{\delta}{F^{(0)\, 2} +\delta^2\gamma_{j}\gamma_e}\left[\gamma_{e}E^{1D}n^{(0)}F^{(0)}+\frac{F^{(0)}}{2}\, {\partial\over
\partial x}\mbox{Im}\, g_{S}\right. \nonumber \\
&&\quad \quad \quad\left. - \frac{\delta\gamma_{e}}{2}\, {\partial\over
\partial x}(n^{(0)}-\mbox{Re}\, g_{S}) -F^{(0)}\mbox{Re}\, h_{S}+\delta
\gamma_{e}\mbox{Im}\, h_{S}\right] =\langle J\rangle_\theta,  \label{eq60}
\end{eqnarray}
where $\langle J\rangle_\theta$ is the total current density averaged over one period of $\theta$. Equations (\ref{eq59}), (\ref{eq60}) and (\ref{eq52}) (the Poisson equation) describe the Bloch oscillations. 

It is important to note that (\ref{eq58}) and (\ref{eq60}) are continuity and Amp\`ere's equations {\em averaged over the fast scale $\theta$}. The total current density depends on the bias condition. For a dc voltage biased SL of nondimensional length $L=(N+1)l/[x]$, we have
\begin{equation}
\frac{1}{L}\int_0^L F(x,t)\, dx = \phi,Ê\label{eq61}
\end{equation}
where $\phi=eV/[\hbar\nu_e (N+1)])$ is a dimensionless average field proportional to the constant applied voltage $V$. Integrating the Amp\`ere's equation (\ref{eq32}) and using $d\phi/dt=0$, we obtain 
\begin{eqnarray}
J= \frac{1}{L}\int_0^L J_n dx =  \frac{1}{L}\int_0^L [J_{n,S}-\mbox{Im }(A e^{-i\theta})]\, dx,  \label{eq62}
\end{eqnarray}
where we have used $f_1\sim f_{1,S}+Ae^{-i\theta}$ and $J_n=-$ Im$f_{1,S}=J_{n,S}-$ Im$(Ae^{-i\theta})$, where $J_{n,S}$ and $\theta$ are given by (\ref{eq39}) and (\ref{eq45}), respectively. Eq. (\ref{eq60}) and the dc voltage bias condition yield
\begin{eqnarray}
\langle J\rangle_\theta =  \frac{1}{L}\int_0^L J_{n,S}\, dx\Longrightarrow J-\langle J\rangle_\theta=-\frac{1}{L}\int_0^L\mbox{Im }(A e^{-i\theta})\, dx.  \label{eq63}
\end{eqnarray}

\subsection{Insufficiency of the KSS-Poisson model}
The local equilibrium (\ref{eq25}) depends only on $n$, therefore $g$ is a function of $n$ and $F$ but it does not depend on $A$ and $\theta$. Thus $g^{(0)}_{-1}=0$, the second term in the right hand side of (\ref{eq59}) is zero and therefore $A(x,t)=A(x,0)\, e^{-(\gamma_e+\gamma_j)t/2}$. The amplitude of the BOs decays exponentially fast to zero. This is consistent with the previous result that the hydrodynamic limit yields only a drift-diffusion equation for $n$ and the electric field \cite{BEP03}.

\section{Dissipative BGK collision model}
\label{sec:4}
We have shown that the KSS-Poisson model cannot sustain BOs because its local equilibrium function does not depend on $J_n$ and $E$ and therefore does not depend on the Bloch phase $\theta$ when (\ref{eq57}) is used. Equation (\ref{eq60}) becomes a drift-diffusion equation in this case. Similarly to the original BGK collision model \cite{bgk}, we need a local equilibrium distribution that depends on $n$, $J_n$ and $E$ in order to obtain a richer set of hydrodynamic equations. To account for thermal effects, we replace the following more general Fermi-Dirac distribution instead of $f^{1D}$ \cite{BC08,BAC11}\cite{footnote2} :
\begin{eqnarray} 
&&f^{1D\alpha}(k;\mu_{\alpha},u_\alpha,T_\alpha) = \frac{m^{*}k_{B}
T_{\alpha}}{\pi\hbar^2}\, \ln\left[1+\exp\left(\frac{\mu_{\alpha} + \hbar k 
u_{\alpha} - {\cal E}(k)}{k_{B}T_{\alpha}}\right)\right]\!,  \label{eq64}
\end{eqnarray}
in dimensional units, or
\begin{equation}
f^{1D\alpha}(k;\tilde{\beta},\tilde{u},\tilde{\mu})=\frac{m^{*}\Delta}{2\pi\tilde{\beta} \hbar^2N_D}\,\ln\left(1+ e^{\tilde{\mu}+\tilde{u}k-\tilde{\beta}+\tilde{\beta} \cos k}\right),   \label{eq65}
\end{equation}
with 
\begin{equation}
\tilde{\mu}= {\mu_{\alpha}\over k_{B}T_{\alpha}},\quad \tilde{u} = {\hbar 
u_{\alpha}\over k_{B}T_{\alpha}l}, \quad\tilde{\beta}= {\Delta\over 2k_{B}
T_{\alpha}},    \label{eq66}
\end{equation}
in nondimensional units. In (\ref{eq64}), $\hbar u_{\alpha} k$ should be considered a periodic function of $k$ with period $2\pi/l$. Then $f^{1D\alpha}$ is $2\pi/l$-periodic in $k$, same as the electron distribution function $f$. The multipliers $\mu_{\alpha}$, $u_{\alpha}$ and $T_{\alpha}$ should be 
selected so that the electron density (\ref{eq4}), the electronic current density (\ref{eq13}) and the mean energy (\ref{eq14}) satisfy the equations:  
\begin{eqnarray} 
&& \frac{ l}{ 2\pi} \int_{-\pi/l}^{\pi/l} f^{1D\alpha}\, dk=n,   \label{eq67}\\
&& \frac{e}{ 2\pi} \int_{-\pi/l}^{\pi/l} v(k)\, f^{1D \alpha}\, dk= (1-\alpha_{j}) J_{n},
\label{eq68}\\
&& \frac{l}{ 2\pi n} \int_{-\pi/l}^{\pi/l}  \left[\frac{\Delta}{ 2} -{\cal E}(k)\right] 
f^{1D \alpha}\, dk = \alpha_{e} E_{0} + (1-\alpha_{e}) E. \label{eq69}
\end{eqnarray} 
Here $\alpha_{j}$ and $\alpha_{e}$ are dimensionless restitution coefficients taking values on the interval $[0,1]$ (see below). $E_{0}$ is the mean energy at the lattice temperature of the global equilibrium reached by a finite SL with insulating contacts and zero voltage bias. The nondimensional versions of (\ref{eq67})-(\ref{eq69}) are 
\begin{eqnarray} 
&& \frac{1}{ 2\pi} \int_{-\pi}^{\pi} f^{1D\alpha}\, dk=n,   \label{eq70}\\
&& \frac{1}{ 2\pi} \int_{-\pi}^{\pi} \sin k\, f^{1D \alpha}\, dk= (1-\alpha_{j}) J_{n},
\label{eq71}\\
&& \frac{1}{ 2\pi n} \int_{-\pi}^{\pi}  \cos k\, f^{1D \alpha}\, dk = \alpha_{e} E_{0} + (1-\alpha_{e}) E. \label{eq72}
\end{eqnarray} 
The nondimensional multipliers $\tilde{\mu}$, $\tilde{u}$ and $\tilde{\beta}$ are functions of $n$, $J_{n}$ and $E$ determined by solving (\ref{eq70})-(\ref{eq72}). With these definitions, the collision operator $Q_e(f)$ of (\ref{eq5}) with $f^{1D\alpha}$ instead of $f^{1D}$ satisfies
\begin{eqnarray}
&& \int_{-\pi/l}^{\pi/l} Q_{e}(f)\, dk = 0,  \label{eq73}\\
&& \frac{e}{2\pi}\,\int_{-\pi/l}^{\pi/l} v(k)\, Q_{e}(f)\, dk = -\nu_{e}\alpha_{j} J_{n},  \label{eq74}\\
&& \frac{l}{2\pi n}\,\int_{-\pi/l}^{\pi/l} \left[\frac{\Delta}{2}-{\cal E}(k)
\right] Q_{e}(f)\, dk = -\nu_{e} \alpha_{e} (E-E_{0}). \label{eq75}
\end{eqnarray} 
In nondimensional units, we have 
\begin{eqnarray}
\frac{1}{2\pi}\,\int_{-\pi}^{\pi} e^{ik} (f-f^{1D\alpha})\, dk = -\alpha_{e}n (E-E_{0})- \alpha_{j} iJ_{n},  \label{eq76}
\end{eqnarray} 
instead of (\ref{eq74}) and (\ref{eq75}), provided we select $\nu=\nu_e$ as our unit of collision frequency. The restitution coefficients $\alpha_{j}$ and $\alpha_{e}$, $0\leq \alpha_{j,e}\leq 1$, measure the fraction of momentum and of energy lost in inelastic collisions, and correspond to the single restitution coefficient used in granular gases \cite{BMD}. Obviously for $\alpha_{e,j}=0$ the collisions are elastic. Note that we do not use the temperature $T_{\alpha}=\alpha T$ as in granular gases because the relation between energy density and temperature is not linear in the present case. To simplify matters, we shall assume that the restitution coefficients are constant. For space independent solutions of the kinetic equation, this leads to exponentially fast decay of the average energy and momentum in contrast with the algebraic decay of energy found in granular gases \cite{BMD}. 

\subsection{Choice of local distribution function}
The distribution function (\ref{eq64}) has the same form as the equilibrium distribution for an electron-phonon collision term in which the phonon distribution is $n_q=(e^{\beta\hbar(\omega_q-\mathbf{q}\cdot \mathbf{u})}-1)^{-1}$, where $\omega_q$ is the phonon frequency corresponding to a wave vector $\mathbf{q}$ and $\mathbf{u}$ is the average velocity. In fact, the electron-phonon collision term for a bulk semiconductor is \cite{reg85}
\begin{eqnarray} 
&& C[f](\mathbf{k}) = \int_{\mathbb{B}} K(|\mathbf{k}-\mathbf{k}'|)\, \{\delta({\cal E}'-{\cal E}+\hbar\omega_{q}-\hbar u\cdot q)\, [ n_{q} f' (1-f) \nonumber\\
&&\quad\quad\quad\, - (1+n_{q})\, f (1-f')]+\, \delta({\cal E}'-{\cal E}-
\hbar\omega_{q}+\hbar\mathbf{q}\cdot\mathbf{u})\nonumber\\
&&\quad\quad\quad\,\times [(1+n_{q})\, f'(1-f) - n_{q} f\,(1-f')]\}\, d\mathbf{k}',\label{eq77}\\
&& n_{q} = \frac{1}{e^{\beta\hbar(\omega_{q}-\mathbf{u}\cdot\mathbf{q})}-1}\,,\quad \mathbf{q}= \mathbf{k}-\mathbf{k}'. \label{eq78}
\end{eqnarray} 
Here $K(|q|)$ depends on the phonon type, $\mathbb{B}$ is the Brillouin zone and $f=f(\mathbf{k})$, $f'=f(\mathbf{k}')$ with similar notation for the dispersion relation $\mathcal{E}(\mathbf{k})$. For a kinetic equation $\partial_t f+ \mathbf{v}(\mathbf{k})\cdot\nabla_xf= C[f](\mathbf{k})$, if $s$ is a function of $\mathbf{k}$ and $f$ with $0\leq f\leq 1$, we find
\begin{eqnarray}  
\frac{\partial}{\partial t}\int_{\mathbb{B}} s(\mathbf{k},f)\, d\mathbf{k} + \nabla_{x}\cdot\int_{\mathbb{B}} \mathbf{v}(\mathbf{k})\, s(\mathbf{k},f)\, d\mathbf{k} = - \int_{\mathbb{B}}\int_{\mathbb{B}} 
K(|\mathbf{k}-\mathbf{k}'|) \nonumber\\
\times \delta({\cal E}'-{\cal E}+\hbar\omega_{q}-\hbar \mathbf{q}\cdot\mathbf{u})\, e^{-\beta ({\cal E}-\hbar \mathbf{k}\cdot\mathbf{u})}  (1+n_{q}) (1-f)\, (1-f')\nonumber\\
\times \left( \frac{e^{\beta ({\cal E}'-\hbar \mathbf{u}\cdot \mathbf{k}')} f'}{1-f'} - \frac{e^{\beta ({\cal E}-\hbar\mathbf{k}\cdot \mathbf{u})} f}{1-f}  \right)  \left(\frac{\partial s}{\partial f'} - \frac{\partial s}{\partial f} \right) d\mathbf{k} d\mathbf{k}'.  \label{eq79}
\end{eqnarray} 
The right hand side of this equation is always less or equal than zero for 
\begin{eqnarray}
\frac{\partial s}{\partial f} &=& \ln\left(\frac{e^{\beta ({\cal E}-\hbar\mathbf{k}\cdot \mathbf{u})} f}{1-f}\right),\quad\mbox{i.e. for an entropy density,} \label{eq80}\\
s(\mathbf{k},f) &=& f\,\ln f + (1-f)\,\ln(1-f) + \beta ({\cal E}-\hbar\mathbf{k}\cdot\mathbf{u}) f, \label{eq81}
\end{eqnarray}
and the corresponding integral over $k$ is a Lyapunov functional for homogeneous distributions. (Note that $s\geq \beta {\cal E} f -1$, and the corresponding integral over $\mathbf{k}$ is bounded below because both the energy and the volume of the Brillouin zone are finite). This is the H-theorem for the electron-phonon kinetic equation. From (\ref{eq82}), we see that the corresponding equilibrium solution satisfies $e^{\beta ({\cal E}-\hbar\mathbf{k}\cdot\mathbf{u})} f/(1-f)=e^{\beta\mu}$ (independent of $\mathbf{k}$), and therefore
\begin{eqnarray}
f_{\rm eq}(\mathbf{k}) = \frac{1}{1+ e^{\beta\, [{\cal E}(\mathbf{k}) - \hbar\mathbf{k}\cdot\mathbf{u}- \mu]} }, \label{eq82}
\end{eqnarray}
 is the equilibrium distribution. If we use (\ref{eq82}) for a SL, then the dispersion relation is ${\cal E}(\mathbf{k})=\mathcal{E}(k)+\hbar^2\mathbf{k}_\perp^2/(2m^*)$, and integration of (\ref{eq82}) over the lateral wave vector $\mathbf{k}_\perp$ yields a distribution function of the same form as (\ref{eq64}) provided $\mathbf{u}$ is directed in the SL growth direction.

A different choice of local distribution function could be 
\begin{eqnarray} 
&&f^{1D\alpha}(k;\mu_{\alpha},k_\alpha,T_\alpha) = \frac{m^{*}k_{B}
T_{\alpha}}{\pi\hbar^2}\, \ln\!\left[1+\exp\left(\frac{\mu_{\alpha} - {\cal E}(k-k_\alpha)}{k_{B}T_{\alpha}}\right)\right]\!,  \label{eq83}
\end{eqnarray}
instead of (\ref{eq64}). The multipliers would now be $\mu_\alpha$, $T_\alpha$ and $k_\alpha$ (instead of $u_\alpha$), to be selected so that conditions (\ref{eq67})-(\ref{eq69}) be satisfied. The choice of $k_\alpha$ would correspond to the wave packet momentum in Lei's formulation \cite{lei95}. For the tight binding dispersion relation (\ref{eq1}), substitution of $\cos(k-k_\alpha)=\cos k\cos k_\alpha+\sin k\sin k_\alpha$ in (\ref{eq83}) would yield a distribution of the following type:
\begin{eqnarray} 
&&f^{1D\alpha}(k;\mu_{\alpha},P_\alpha,T_\alpha) = \frac{m^{*}k_{B}
T_{\alpha}}{\pi\hbar^2}\, \ln\!\left[1+\exp\left(\frac{\mu_{\alpha} - {\cal E}(k)-v(k) P_\alpha}{k_{B}T_{\alpha}}\right)\right]\!.  \label{eq84}
\end{eqnarray}
The distribution (\ref{eq84}) could have been obtained from the maximum entropy principle as suggested in \cite{wu94} and, for the tight binding dispersion relation, it becomes (\ref{eq83}) selecting appropriately the multipliers $\mu_\alpha$ and $T_\alpha$; cf.\ \ref{app0}. We show in \ref{app0} that the BO solutions corresponding to transport equations that have local equilibrium distribution functions (\ref{eq83}) and (\ref{eq84}) (with tight binding dispersion relation) are always damped. 

\subsection{Equations of the model}
Since (\ref{eq74}) and (\ref{eq75}) show that our collision model dissipates both momentum and energy, we propose a simpler equation for the distribution function with $Q_p(f)=0$ in (\ref{eq2}) and (\ref{eq64}) as the local distribution function instead of (\ref{eq6}). Recapitulating, the equations governing our inelastic BGK model are (\ref{eq2}) and (\ref{eq3}) with $Q_p=0$ and $Q_{e}=-\nu_{e}(f-f^{1D\alpha})$ given by (\ref{eq64}) and (\ref{eq67})-(\ref{eq69}). $n$, $J_{n}$ and $E$ are the moments of the distribution function given by (\ref{eq4}), (\ref{eq13}) and (\ref{eq14}), respectively. In nondimensional units, the equations of the model are (\ref{eq22})-(\ref{eq24}) with $\nu_p=0$, $\nu=\nu_e$ and $f^{1D\alpha}$ given by (\ref{eq65}) instead of $f^{1D}$: 
\begin{eqnarray} 
&&F\,\frac{\partial f}{\partial k}+ f-f^{1D\alpha}=-\delta\left(\frac{\partial f}{\partial t} + \sin k\, \frac{\partial f}{ \partial x}\right)\!,  \label{eq85}\\
&&\frac{\partial F}{\partial x} = n-1,  \label{eq86}\\   
&& n = \frac{1}{2\pi} \int_{-\pi}^{\pi} f(x,k,t)\, dk= \frac{1}{2\pi} \int_{-\pi}^{\pi} f^{1D\alpha}(x,k,t)\, dk,    \label{eq87}\\
&&\frac{1}{2\pi}\,\int_{-\pi}^{\pi} e^{ik} (f-f^{1D\alpha})\, dk = -\delta\gamma_{e}n(E-E_{0})- \delta\gamma_{j} iJ_{n},  \label{eq88}\\
&& f^{1D\alpha}(k;n,E,J_n) = \frac{m^{*}\Delta}{2\pi\tilde{\beta}\hbar^2N_D}\, \ln\left(1+ e^{\tilde{\mu}+\tilde{u}k-\tilde{\beta}+\tilde{\beta}\cos k}\right)\!.  \label{eq89}
\end{eqnarray} 
Here we have assumed that the restitution coefficients are of order $\delta$ and defined $\gamma_e$ and $\gamma_j$ for this model as: 
\begin{equation}
\alpha_{e}=\delta\gamma_{e},\quad \alpha_j=\delta\gamma_j, \label{eq90}
\end{equation} 
instead of (\ref{eq29}) for the KSS equation. The restitution coefficients $\alpha_{e,j}$ can be fitted to experimentally obtained current--voltage characteristic curves in the same way as the KSS collision frequencies $\nu_e$ and $\nu_p$. We find $\nu\alpha_e=\nu_e$ and $\nu\alpha_j=\nu_e+\nu_p$. When modeling a finite SL, we need boundary conditions for $f$ and $F$ at the contacts attached to the SL boundaries and an initial condition for $f$. See References \cite{BGr05} and \cite{elena} for a discussion. 

\subsection{Boltzmann distribution}
We can simplify the previous formulas in the low temperature limit in which $\tilde{\beta}\to\infty$, $\tilde{u}=O(\tilde{\beta})$, $\tilde{\mu}\to - \infty$ in (\ref{eq89}), which becomes the Boltzmann distribution
\begin{eqnarray} 
f^B = \frac{m^*\Delta }{ 2\pi\hbar^2\tilde{\beta}N_D}\,  e^{\tilde{\mu} + \tilde{u}
k -\tilde{\beta}\, (1-\cos k)}, \label{eq91}
\end{eqnarray} 
and integrals over $k$ are calculated using Laplace's method. For sufficiently high temperature, the Boltzmann distribution (\ref{eq91}) is again a good approximation and it yields simpler formulas. The parameter $\tilde{\mu}$ can be explicitly calculated using (\ref{eq91}) in (\ref{eq87}) and the resulting distribution is 
\begin{eqnarray} 
f^B = n\,\frac{\pi\, e^{ \tilde{u}k +\tilde{\beta}\cos k}}{\int_{0}^\pi 
e^{\tilde{\beta}\cos K}\cosh(\tilde{u}K)\, dK},   \label{eq92}
\end{eqnarray} 
in which $\tilde{u}$ and $\tilde{\beta}$ are obtained in terms of $J_{n}/n$ and $E$ by solving (\ref{eq88}). As shown in \ref{app1}, the latter equations yield 
\begin{eqnarray} 
&&\frac{(1-\alpha_{j})\, J_{n}}{ n} =\frac{\tilde{u}}{\tilde{\beta}}-\frac{ e^{- \tilde{\beta}}\sinh(\tilde{u}\pi)}{\tilde{\beta}\int_{0}^\pi e^{\tilde{\beta}\cos K}\cosh(\tilde{u}K)\, dK}\nonumber\\
&&\quad\quad\quad\quad\quad=\frac{\tilde{u}}{\tilde{\beta}}\left[1 - \frac{e^{-\tilde{\beta}}}{I_0(\tilde{\beta})+2\tilde{u}^2\sum_{l=1}^\infty \frac{(-1)^l}{\tilde{u}^2+l^2}I_l(\tilde{\beta})}\right]\!,   \label{eq93}\\
&&E-\alpha_{e}(E-E_0) = \frac{ \int_{0}^\pi e^{\tilde{\beta}\cos K} \cos K\cosh(\tilde{u}K)\, dK}{\int_{0}^\pi e^{\tilde{\beta}\cos K}\cosh(\tilde{u}K)\, dK}\nonumber\\
&&\quad\quad\quad\quad\quad\quad\quad = \frac{I_1(\tilde{\beta})+\tilde{u}^2\sum_{l=1}^\infty \frac{(-1)^l}{\tilde{u}^2+l^2}[I_{l-1}(\tilde{\beta})+I_{l+1}(\tilde{\beta})]}{I_0(\tilde{\beta})+2\tilde{u}^2\sum_{l=1}^\infty \frac{(-1)^l}{\tilde{u}^2+l^2}I_l(\tilde{\beta})}, \label{eq94}
\end{eqnarray} 
where $I_{s}(x)$ are modified Bessel functions \cite{AS}. At the lattice temperature, $\tilde{\beta}_{0}=\Delta/(2k_{B}T_{0})$, and for zero current, $\tilde{u}=0$, $E=E_{0}$, and (\ref{eq94}) yields
\begin{eqnarray} 
E_0=\frac{I_{1}(\tilde{\beta}_{0})}{I_{0}(\tilde{\beta}_{0})}.  \label{eq95}
\end{eqnarray} 
Further simplification follows if we impose $\alpha_{j}=1$ in (\ref{eq93}) (which implies $\tilde{u}=0$) so that the BGK collision term dissipates momentum and energy according to (\ref{eq88}). Then (\ref{eq92}) becomes
\begin{eqnarray} 
f^B = \frac{n\, e^{\tilde{\beta}\cos k}}{I_{0}(\tilde{\beta})},   \label{eq96}
\end{eqnarray} 
and $\tilde{\beta}$ is obtained in terms of $E$ by solving (\ref{eq88}) with $\tilde{u}=0$, i.e. 
\begin{eqnarray} 
\alpha_{e}E_{0}+(1-\alpha_{e})E = \frac{ I_{1}(\tilde{\beta})}{I_{0}(\tilde{\beta})}. \label{eq97}
\end{eqnarray} 
The Fourier coefficients of the Boltzmann distribution (\ref{eq96}) are simply
\begin{eqnarray} 
f^B_{j} = \frac{1}{2\pi}\int_{-\pi}^{\pi} e^{-ijk}f^B(k;n)\, dk= n\,
\frac{ I_{j}(\tilde{\beta})}{I_{0}(\tilde{\beta})}. \label{eq98}
\end{eqnarray} 

\section{Chapman-Enskog method for almost elastic collisions}
\label{app2}
In this Section, we shall derive modulation equations for $n$, $F$ and $A$ in the case of almost elastic collisions with $0<\alpha_{e,j}\ll 1$, more precisely in the double limit of small ``Knudsen'' number $\delta$ and vanishing restitution coefficients $\alpha_{e,j}=\delta\gamma_{e,j}$. For the case of granular gases, Sela and Goldhirsch derived hydrodynamic equations from the inelastic Boltzmann equation using a CEM in a similar double limit \cite{sel98}. We shall use the CEM \cite{BT10} to obtain modulation equations for the electric field, the electron density and the amplitude $A$ of the BOs. Then we will compare these equations with (\ref{eq59}) and (\ref{eq60}) and identify $g$ and $h$. 

We can repeat the calculations of Section \ref{sec:3} with the local equilibrium distribution (\ref{eq89}) and get the same modulation equations (\ref{eq59})-(\ref{eq60}) except that $E_0$ replaces $E^{1D}$ in them and $\gamma_{e,j}$ are defined by (\ref{eq90}) instead of (\ref{eq29}). Now the local equilibrium distribution $f^{1D\alpha}$ depends on $f_1=nE-iJ_n$ and therefore it depends on the Bloch phase $\theta$ through (\ref{eq57}). Then the second term in the RHS of (\ref{eq59}) is no longer zero and BOs are not necessarily damped by scattering. This term will be now identified by using the Chapman-Enskog method to derive modulation equations that will be matched to (\ref{eq58})-(\ref{eq60}). To implement the CEM, we assume that the distribution function $f$ is a function of $\theta$, $k$, $x$ and $t$, which is $2\pi$-periodic in $\theta$ and in $k$ and that $F$ is of order 1. Eq.\ (\ref{eq85}) becomes
\begin{eqnarray} 
&& \mathcal{L}f - f^{1D\alpha} = -\delta\left(\frac{\partial f}{\partial t} + 
\sin k\, {\partial f\over\partial x}\right),  
\label{eq99}\\
&&\mathcal{L}u(k,\theta)= F \left({\partial\over \partial\theta}+{\partial\over 
\partial k}\right)\!u(k,\theta)+ u(k,\theta), \label{eq100}
\end{eqnarray}
Equations (\ref{eq99})-(\ref{eq100}) with $F=O(1)$ display a dominant balance between the collisions, the force due to the electric field and the change of $f$ over the fast time scale $\theta$. We are ignoring a possible fast relaxation from an initial distribution function to the BO distribution $f(x,k,\theta,t;\delta)$ that depends only on one fast time scale $\theta$. We are imposing the condition that $f$ be periodic in $\theta$ and considering the possibility of slow modulations of the BOs in the time scale $t$.

The moment $f_{1}=f_{1S}+\Phi$, $\Phi=A e^{-i\theta}+O(\delta)$, has a dominant part of order one, $Ae^{-i\theta}$, and a remainder of order $\delta$. The remainder vanishes as $\delta\to 0$ and it can be chosen not to contain a term proportional to $e^{-i\theta}$. Thus we assume:
\begin{equation}
f_{1}= A e^{-i\theta} + \delta B + \delta^2 C + O(\delta^3). \label{eq101}
\end{equation}
The local equilibrium $f^{1D\alpha}$ is a function of $k$, $n$ and $f_{1}$ through (\ref{eq87}) - (\ref{eq89}). Due to (\ref{eq101}), $f^{1D\alpha}$ is a $2\pi$-periodic function of $k$ and of $\theta$, which also depends on the slowly-varying functions $n(x,t)$, $F(x,t)$, $A(x,t)$, $B(x,t)$ and $C(x,t)$. 

Using the Fourier series
\begin{eqnarray} 
&&f^{1D\alpha}(k,\theta;\delta)=\sum_{j,l}f^{1D\alpha}_{j,l}e^{i(jk+l\theta)},
\label{eq102}\\
&&f(k,\theta;x,t,\delta)=\sum_{j,l}f_{j,l}(x,t;\delta) e^{ijk+il\theta}\!,\, f_{j,l} = \int_{-\pi}^\pi\! \int_{-\pi}^\pi \! e^{-ijk-il\theta}f\frac{dk d\theta}{(2\pi)^2} , \label{eq103}
\end{eqnarray}
in (\ref{eq99}) with $\delta=0$, we immediately obtain
\begin{equation}
f^{(0)}(k,\theta;x,t)=\sum_{j,l}f_{j,l}^{(0)}(x,t)\, e^{i(jk+l\theta)}, \quad f^{(0)}_{j,l} = \frac{f^{1D\alpha(0)}_{j,l}}{1+iF(j+l)},  \label{eq104}
\end{equation}
where the superscripts (0) refer to having set $\delta=0$ in (\ref{eq99}) (see below).

The CEM starts from a leading order expression for the distribution function, (\ref{eq104}), which does not depend explicitly on $x$ and $t$. Instead, it depends on $k$ and $\theta$, and it is a function of quantities that vary slowly with $x$ and $t$ ($n$, $F$, $A$, $B$, $C$, \ldots). $B$, $C$, \ldots are to be calculated in terms of $A$, $n$, $F$ and their differentials. While $A$, $n$ and $F$ are not expanded in powers of $\delta$,  their partial derivatives with respect to time (and therefore their equations of motion) are expanded instead. Thus the Chapman-Enskog Ansatz is
\begin{eqnarray} 
&& f(x,k,t;\delta) = \sum_{m=0}^{\infty} f^{(m)}(k,\theta;F,n,A,B,C)\, 
\delta^{m} , \label{eq105}\\
&& \frac{\partial F}{\partial t}+\sum_{m=0}^\infty {\cal J}^{(m)}(F,n,A,B,C)\,
\delta^{m}=\langle J\rangle_\theta,  \label{eq106}\\
&& \frac{\partial n}{\partial t} = -\sum_{m=0}^{\infty} \frac{\partial}{\partial x} \mathcal{J}^{(m)}(F,n,A,B,C)\,\delta^{m} ,   \label{eq107}\\
&&\frac{\partial A}{\partial t}=\sum_{m=0}^{\infty}\mathcal{A}^{(m)}(F,n,
A,B,C)\,\delta^{m}.   \label{eq108}
\end{eqnarray} 
We have used the Poisson equation (\ref{eq86}) and (\ref{eq106}) to obtain (\ref{eq107}). The local distribution function $f^{1D\alpha}$ can be expanded in powers of $\delta$,
\begin{eqnarray} 
f^{1D\alpha}= \sum_{m=0}^\infty f^{1D\alpha(m)}\delta^m, \label{eq109}
\end{eqnarray}
and then (\ref{eq87}) - (\ref{eq88}) yield the following compatibility
conditions:
\begin{eqnarray} 
&&f_{0,0}^{(m)}=f^{1D\alpha(m)}_{0,0}= n\,\delta_{0m}, \label{eq110}\\
&& f_{1,-1}^{(m)}=f^{1D\alpha(m)}_{1,-1}=A\,\delta_{0m}, \label{eq111}\\
&& f^{(1)}_{1,0}= B,\quad f^{1D\alpha(1)}_{1,0}= B+\gamma_{e}n\, E_{0}, \label{eq112}\\
&& f^{(2)}_{1,0}= C, \quad f^{1D\alpha(2)}_{1}= C-\gamma_{e}n\,\mbox{Re} B - i \gamma_{j} \mbox{Im} B,   \label{eq113}
\end{eqnarray} 
and so on. Inserting (\ref{eq105}) - (\ref{eq109}) into (\ref{eq99}), we obtain a hierarchy of linear equations for the $f^{(m)}$ whose right hand sides contain the functionals ${\cal J}^{(m)}$ and ${\cal A}^{(m)}$. The latter are calculated in such a way that the compatibility conditions (\ref{eq110}) - (\ref{eq113}) hold.

The equations for $f^{(1)}$ and $f^{(2)}$ are:
\begin{eqnarray} 
&&\mathcal{L} f^{(1)} -f^{1D\alpha(1)}=  \left. - \left(\frac{\partial f^{(0)}}{\partial t}\right|_{0} + \sin k\, \frac{\partial f^{(0)}}{\partial x}\right)\!,  \label{eq114}\\
&&\mathcal{L} f^{(2)}-f^{1D\alpha(2)} = \left. -\left(\frac{\partial f^{(1)}}{\partial t} \right|_{0} + \sin k\,\frac{\partial f^{(1)}}{\partial x}\right) - \left. \frac{\partial f^{(0)}}{\partial t}\right|_{1} . \label{eq115}
\end{eqnarray} 
The subscripts $m=0,1$ in the right hand side of these equations mean that $\partial F/\partial t$, $\partial n/\partial t$ and $\partial A/\partial t$ are replaced by $(\langle J\rangle_\theta\delta_{m0}-\mathcal{J}^{(m)})$, $-\partial\mathcal{J}^{(m)}/\partial x$ and $-\mathcal{A}^{(m)}$, respectively. 

Upon insertion of (\ref{eq104}) in (\ref{eq114}), the compatibility conditions (\ref{eq110}) - (\ref{eq113}) yield
\begin{eqnarray} 
&&\mathcal{J}^{(0)} = 0,   \label{eq116}\\
&&\mathcal{A}^{(0)} = -\frac{1}{2}(\gamma_{e}+\gamma_{j})\, A
+\frac{1}{2i}\,\frac{\partial f^{(0)}_{2,-1}}{\partial x},  \label{eq117}\\
&& B= \frac{\gamma_{e}nE_{0}}{iF} + \frac{1}{2F}\,\frac{\partial}{\partial 
x}\left(n-\frac{f^{1D\alpha(0)}_{2,0}}{1+i2F}\right).\label{eq118}
\end{eqnarray}
Note that $B$ becomes singular in the limit as $F\to 0$. This is not surprising: we have assumed in this Section that $F=O(1)$ as $\delta\to 0$ so that $\theta\neq 0$ and the Fourier series in $\theta$ of $f_1$ in (\ref{eq101}) has a first harmonic $A e^{-i\theta}$ that is different from all other ones contained in $B$, $C$, etc. If $F$ tends to 0, then the first two terms in (\ref{eq100}) are smaller than the third one and the assumption (\ref{eq101}) breaks down. Despite this shortcoming, we shall use the CEM to identify the closure functions $g$ and $h$ introduced in the previous Section. The coefficients in the resulting modulation equations are no longer singular.

The compatibility conditions (\ref{eq113}) for $f^{(2)}$ provide the following functionals
\begin{eqnarray} 
&&\mathcal{J}^{(1)} = -\mbox{Im }B,   \label{eq119}\\
&&\mathcal{A}^{(1)} =\frac{1}{2i}\,\frac{\partial f^{(1)}_{2,-1}}{\partial x}.  
\label{eq120}
\end{eqnarray}
Then the $\theta$-averaged Amp\`ere's law and the equation for $A$ including up to $O(\delta)$ terms are
\begin{eqnarray} 
&& \frac{\partial F}{\partial t}-\delta\,\mbox{Im}\, B = \langle J\rangle_\theta,\label{eq121}\\
&&\frac{\partial A}{\partial t}= -\frac{1}{2}(\gamma_{e}+\gamma_{j})\, A+\frac{1}{2i}\,\frac{\partial}{\partial x}( f^{(0)}_{2,-1}+\delta\, f^{(1)}_{2,-1}),   \label{eq122}
\end{eqnarray} 
in which $B$ is given by (\ref{eq118}) and the Fourier coefficients of the solution of (\ref{eq114}) are
\begin{eqnarray} 
&& f^{(1)}_{j,l} = \frac{r^{(1)}_{j,l}}{1+iF(j+l)},  \label{eq123}\\
&& r^{(1)} = f^{1D\alpha(1)}-\left(\left.\frac{\partial}{\partial t}\right|_{0}
+\sin k\,\frac{\partial}{\partial x}\right) f^{(0)}.\label{eq124}
\end{eqnarray} 

\subsection{Identification of $g$ and $h_S$ in the modulation equations}
We now impose that Equations (\ref{eq121}) and (\ref{eq122}) match (\ref{eq60}) (with $E_0$ instead of $E^{1D}$) and (\ref{eq59}), respectively. The result is that these equations match term by term in the overlap region 
$$\delta\ll F\ll 1,
$$
(with $B \delta\sim f_{1,S}$) provided
\begin{eqnarray}
&& g_{S}= f^{(0)}_{2,0}, \label{eq125}\\
&& g_{-1}= f^{(0)}_{2,-1} + \delta\,  f^{(1)}_{2,-1}. \label{eq126}
\end{eqnarray}
Both equations hold if 
\begin{eqnarray}
g = f^{(0)}_{2} + \delta\,  f^{(1)}_{2}. \label{eq127}
\end{eqnarray}
We have not yet calculated $h_S$ in (\ref{eq60}). To determine it, we require that the resulting equation for the field coincide with the drift-diffusion equation (\ref{b.36}) derived in \ref{sec:5} for the case of inelastic collisions (without BOs). As seen in \ref{sec:5}, $h_S$ in (\ref{eq60}) should be replaced by the uniform part of $\partial f_{1,S}/\partial t$ in (\ref{eq38}), i.e., 
\begin{eqnarray}
h_S&=&\left.\frac{\partial}{\partial t}\left(\frac{\delta\gamma_{e}nE_{0}(\delta
\gamma_{j}-iF)}{\delta^2\gamma_{e}\gamma_{j}+ F^2}\right)\right|_{0}=- \frac{\delta\gamma_{e}E_{0}(\delta\gamma_{j}-iF)}{\delta^2\gamma_{e}\gamma_{j}+ F^2}\, \frac{\partial J_{n,Su}}{\partial x}\nonumber\\
&+&(\langle J\rangle_\theta-J_{n,Su})\,\frac{\partial}{\partial F}\left(\frac{\delta\gamma_{e}nE_{0}(\delta\gamma_{j}-iF)}{\delta^2\gamma_{e}\gamma_{j}+ F^2}\right). \label{eq128}
\end{eqnarray}
Here $J_{n,Su}= \delta\gamma_e nE_0F/(\delta^2 \gamma_j\gamma_{e}+F^2)$ according to (\ref{eq41}). 

\subsection{Modulation equations}
After straightforward calculations, we obtain the following reduced equations:
\begin{eqnarray} 
&&\frac{\partial F}{\partial t} +\frac{\delta}{F^2+\delta^2\gamma_{j}\gamma_e} \left[ \gamma_{e}E_{0}nF+\frac{F}{2}\, \frac{\partial}{\partial x}\mbox{Im}\,\frac{f^{1D\alpha(0)}_{2,0}}{1+2iF}\right.\nonumber\\
&& \quad\left. - \frac{\delta\gamma_{e}}{2}\,\frac{\partial}{\partial x}\left(n-\mbox{Re}\,\frac{f^{1D\alpha(0)}_{2,0}}{1+2iF}\right) - F \mbox{Re}\, h_{S}+\delta
\gamma_{e}\mbox{Im}\, h_{S}\right] =\langle J\rangle_\theta, \label{eq129}\\
&&\frac{\partial A}{\partial t} =-\frac{\gamma_{e}+\gamma_{j}}{2}
\,A + \frac{1}{2i}\frac{\partial}{\partial x}\!\left(\frac{f^{1D\alpha(0)}_{2,-1}
+\delta\, r^{(1)}_{2,-1}}{1+iF}\right)\!,  \label{eq130}\\
&&r^{(1)}_{2,-1} = f^{1D\alpha(1)}_{2,-1}\! -\!\left(\mathcal{A}^{(0)}
\frac{\partial}{\partial A}+(\langle J\rangle_\theta-J_{n,Su})\frac{\partial}{\partial F}-\frac{\partial J_{n,Su}}{\partial x}\frac{\partial}{\partial n}\right)\!\frac{f^{1D
\alpha(0)}_{2,-1}}{1+iF}\nonumber\\
&&\quad\quad -\frac{1}{2i}\frac{\partial}{\partial x}\!\left(A -\frac{f^{1D\alpha(0)}_{3,-1}}{1+2iF}\right)\!, \label{eq131}
\end{eqnarray}
in addition to (\ref{eq128}) and to the Poisson equation (\ref{eq86}):
\begin{eqnarray}
\frac{\partial F}{\partial x} = n-1.  \label{eq132}
\end{eqnarray} 
To calculate $f^{1D\alpha(1)}_{2,-1}$ in (\ref{eq122}), we need to use
\begin{eqnarray}
f^{1D\alpha(1)} = \left(\tilde{\mu}^{(1)}\frac{\partial}{\partial
\tilde{\mu}^{(0)}}+\tilde{u}^{(1)}\frac{\partial}{\partial\tilde{u}^{(0)}}+
\tilde{\beta}^{(1)}\frac{\partial}{\partial\tilde{\beta}^{(0)}}\right)
f^{1D\alpha(0)}.\label{eq133}
\end{eqnarray}
In this expression, we should substitute $\tilde{\mu}^{(0)}$, $\tilde{u}^{(0)}$ and 
$\tilde{\beta}^{(0)}$ given by simultaneously solving
\begin{eqnarray}
f^{1D\alpha(0)}_{0}=n,\quad f^{1D\alpha(0)}_{1}=A\, e^{-i\theta},\label{eq134}
\end{eqnarray}
and also the solutions $\tilde{\mu}^{(1)}$, $\tilde{u}^{(1)}$, $\tilde{\beta}^{(1)}$ 
of
\begin{eqnarray}
&& f^{1D\alpha(1)}_{0}=0,\label{eq135}\\
&& f^{1D\alpha(1)}_{1}=\gamma_{e}n E_{0}+f_{1,S} -\frac{\gamma_{e}+\gamma_{j}}{2}\, A\, e^{-i\theta}-\frac{\gamma_{e}-\gamma_{j}}{2}\,\overline{A} e^{i\theta}. \label{eq136}
\end{eqnarray}
When we substitute these solutions, (\ref{eq133}) becomes a function of $k$, $\theta$, $n$, $A$ and $f_{1,S}\sim B\delta$ which is $2\pi$-periodic in $k$ and $\theta$. Its Fourier coefficient $f^{1D\alpha(1)}_{2,-1}$ is then inserted in (\ref{eq131}).

\section{Numerical solutions of the modulation equations for $n$, $F$ and $A$}
\label{sec:6}
\subsection{Modulation equations}
According to the results in the previous section, the modulation equations for the local equilibrium distribution $f^{1D\alpha}$ are (\ref{eq129}), (\ref{eq130}) and (\ref{eq132}). The coefficient functions appearing in these equations are given by (\ref{eq128}), (\ref{eq131}) and (\ref{eq133})-(\ref{eq136}). The modulation equations with $r^{(1)}_{2,-1}=0$ and lattice temperature of 300 K were shown in Ref. \cite{BAC11} to exhibit BOs confined to a portion of the SL provided $(\gamma_e+\gamma_j)/2<\gamma_c$ ($\gamma_c$ is a critical value). 

\subsection{Boundary, bias and initial conditions}
The boundary and bias conditions are:
\begin{eqnarray}
&&\langle J\rangle_\theta-\frac{\partial F}{\partial t} =\sigma_0 F \quad\mbox{(at $x=0$),} \quad
\langle J\rangle_\theta-\frac{\partial F}{\partial t} =\sigma_1 n F\quad \mbox{(at $x=L$),}
\label{eq137}\\
&&A= 0, \quad\mbox{at $x=0$ and at $x=L$,}\label{eq138}\\
&& \frac{1}{L}\int_0^L F(x,t)\, dx = \phi. \label{eq139}
\end{eqnarray}
The last equation is the nondimensional dc voltage bias condition (\ref{eq61}) with $\phi= eV/[\hbar\nu_e(N+1)]$. It is also possible to set $\partial A/\partial x=0$ in the contacts and the numerical results are similar. According to (\ref{eq62}) and (\ref{eq63}), the total current density is
\begin{eqnarray}
J&=& \frac{1}{L}\int_0^L J_n dx =  \frac{1}{L}\int_0^L [J_{n,S}-\mbox{Im }(A e^{-i\theta})]\, dx\nonumber\\
&=& \langle J\rangle_\theta-\frac{1}{L}\int_0^L\mbox{Im }(A e^{-i\theta})\, dx, 
\label{eq140}
\end{eqnarray}
for dc voltage bias. In our numerical solutions, we have adopted uniform profiles for $F(x,0)$ and $A(x,0)$ as initial conditions. 

\subsection{Numerical results}
We shall illustrate our results with numerical solutions of (\ref{eq129})-(\ref{eq132}) with the Boltzmann local distribution function (\ref{eq92})-(\ref{eq94}). Using the more general Fermi-Dirac local equilibrium (\ref{eq65})  complicates the numerical procedure by having to calculate one more multiplier at each time step: $\tilde{\mu}$ in addition to $\tilde{\beta}$ and $\tilde{u}$. We shall use indicative values similar to those in Ref.\ \cite{sch98}: $l=5.06$ nm, $\Delta= 16$ meV, $\nu_{e}=10^{14}$ Hz, with $\alpha_{e}=\alpha_{j}=0.006$ so that $\nu_{e}\alpha_{e}= \nu_e\alpha_j= 6\times 10^{11}$ Hz.\cite{footnote3} The 3D doping density is $N_{3D}= 8\times 10^{16}$ cm$^{-3}$, so that $N_{D}=N_{3D}l=4.048\times 10^{10}$ cm$^{-2}$, and $\varepsilon=12.85\,\varepsilon_{0}$. We find $\delta\approx 0.0053$ and $\gamma_{e,j}=\alpha_{e,j}/\delta=1.1269$. We consider a 50-period ($N=50$) dc voltage biased SL with lattice temperature 70 K. We have used contact conductivities $\sigma_{0}=60.6$ $(\Omega\,\mbox{m})^{-1}$ and $\sigma_{1}=15.15$ $(\Omega\,\mbox{m})^{-1}$ which yield dimensionless conductivities $\sigma_{0}=1$ and $\sigma_1=0.25$ in (\ref{eq137}) (conductivity units are $[\sigma]= e^2N_D\Delta l/(2\hbar^2\nu_e)$). Initially, the profiles of $A$ and $F$ are uniform, with a common value 0.5501. 

For $V=0.133$ V (therefore $\phi=0.04$) and after a short transient that depends on the initial conditions, we observe coexisting BOs with frequency about 0.4 THz and Gunn type oscillations with frequency about 14 GHz. See the movie in the Supplementary matterial \cite{movie}. Fig. \ref{fig1} shows several snapshots of the field and $|A|$ profiles of the Gunn type oscillation. While the amplitude of Gunn-type current oscillation is about 0.03 in nondimensional units (as seen in Fig. \ref{fig1}(a) for the total current density averaged over the BOs), the BO part of the current oscillation has a larger amplitude of about 0.2; see Fig. \ref{fig2}(a). Fig. \ref{fig2} illustrates the total current density (\ref{eq140}) of the coexisting 360 GHz Bloch and 13.8 GHz Gunn type oscillations, respectively. For each lattice temperature, there is a critical curve in the plane of restitution coefficients such that, for $(\gamma_e+\gamma_j)/2>\gamma_{\rm crit}$, BOs disappear after a relaxation time but they persist for smaller values of $(\gamma_e+\gamma_j)$ \cite{BAC11}. 
\begin{figure}
\begin{center}
\includegraphics[width=8cm,angle=0]{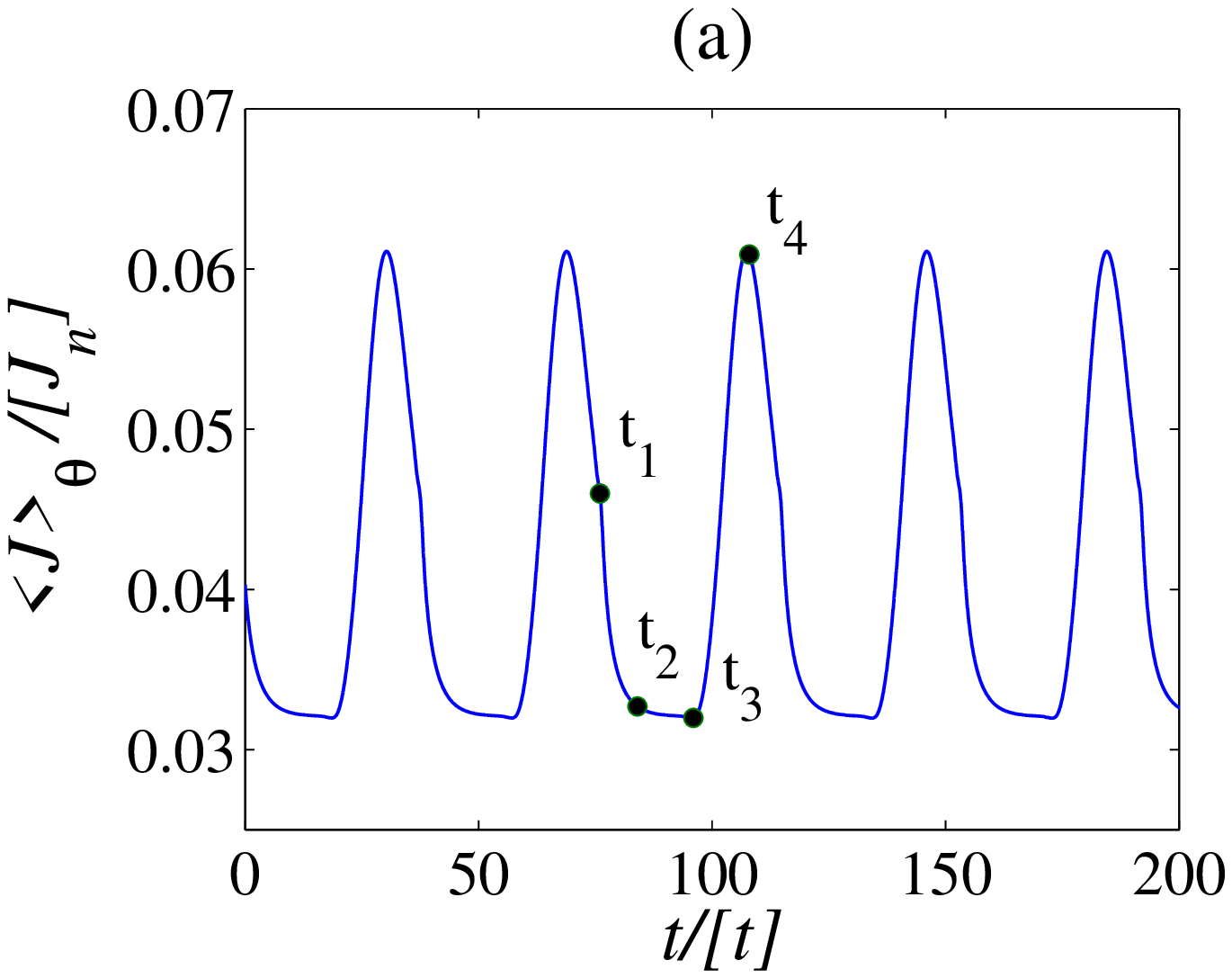}\\
\includegraphics[width=8cm,angle=0]{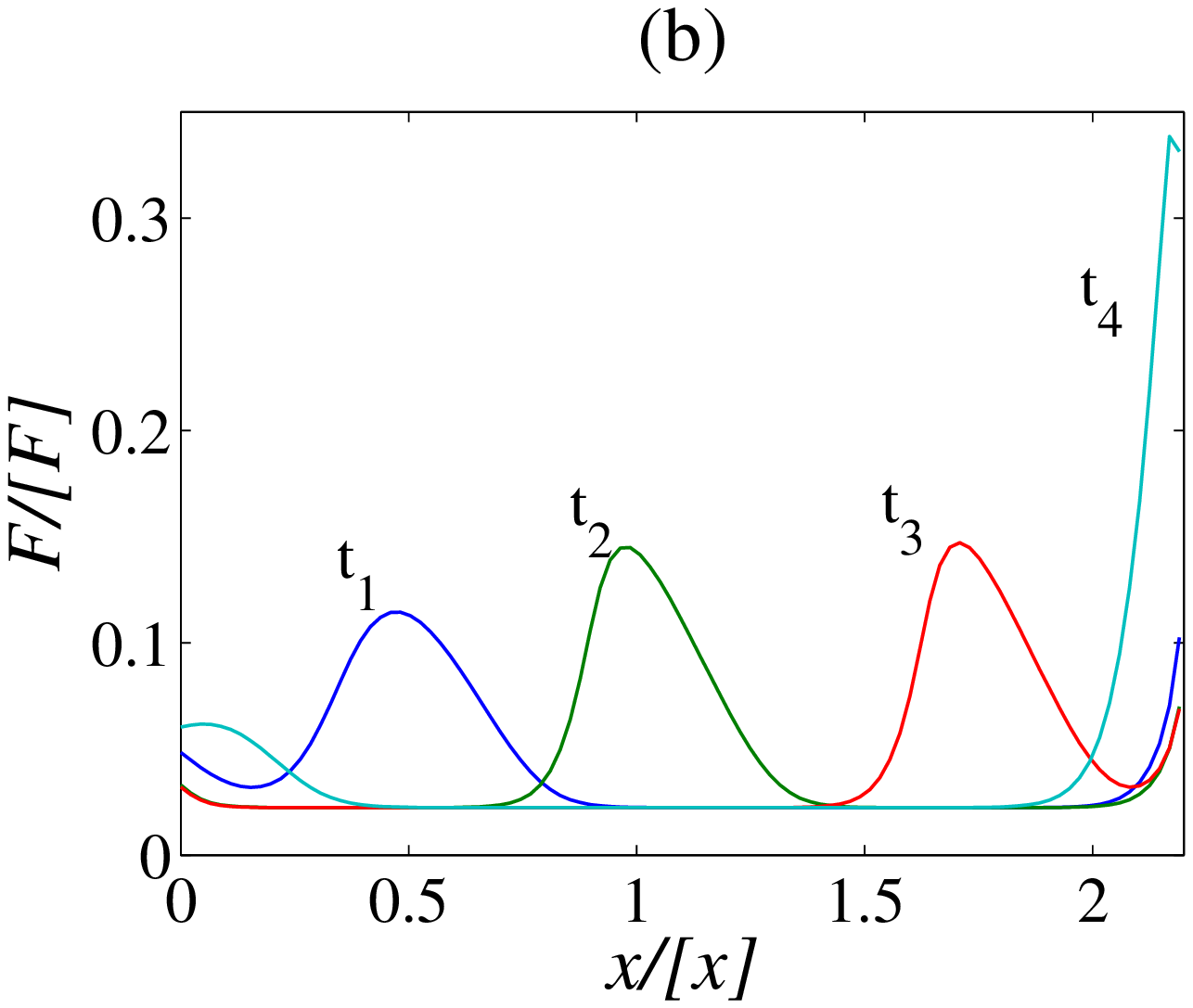}\\
\includegraphics[width=8cm,angle=0]{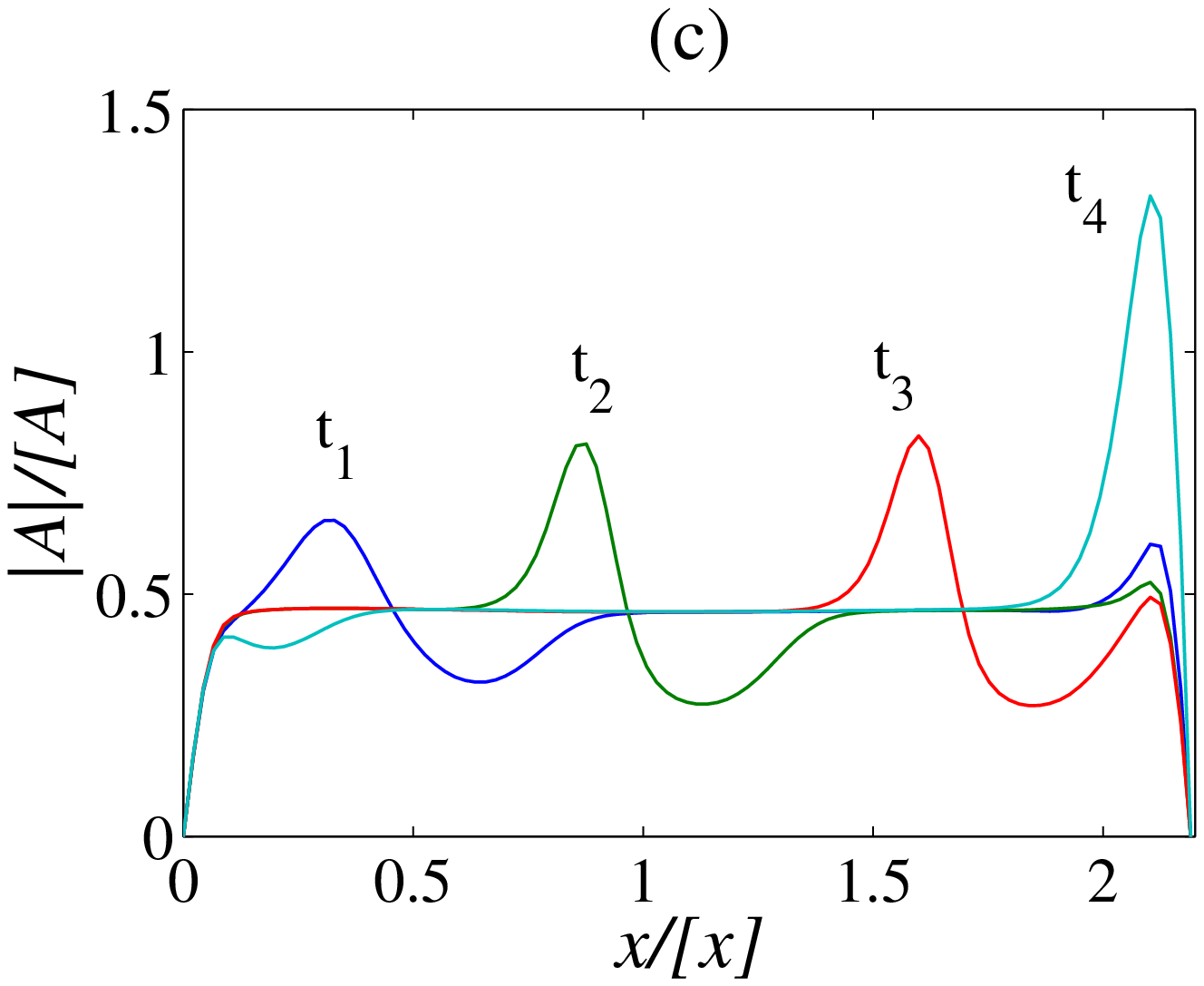}\\
\end{center}
\begin{center}
\vspace{0.2cm} \caption{(a) $\theta$-averaged total current density vs time during coexisting Bloch and Gunn type oscillations at 70 K. (b) Field profile vs space at the times $t_1$ to $t_4$ marked in (a).  (c) Same for the complex BO amplitude profile. To transform the magnitudes in this figure to dimensional units, use Table \ref{t1}. $[A]= N_D$. } \label{fig1}
\end{center}
\end{figure}

\begin{figure}
\begin{center}
\includegraphics[width=8cm,angle=0]{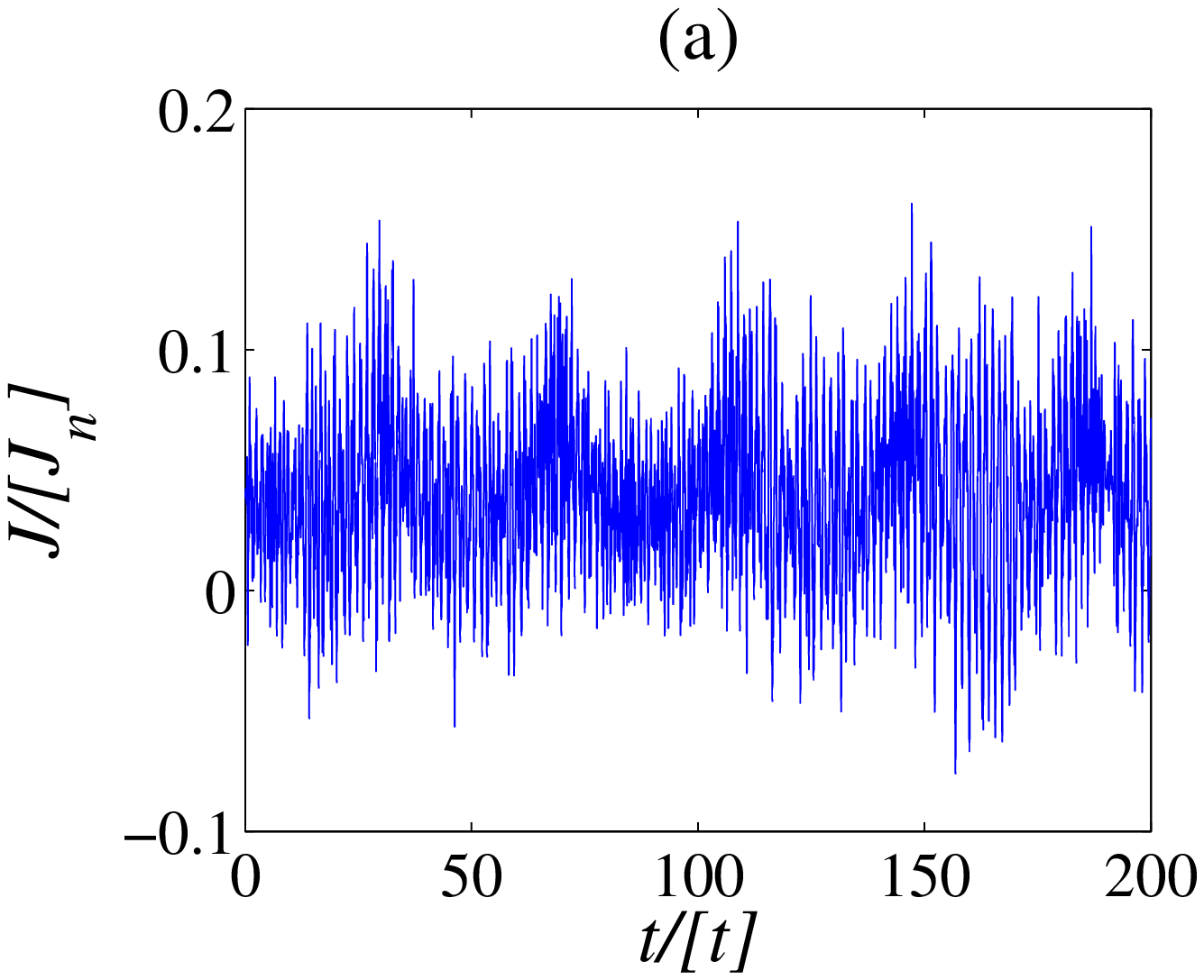}\\
\includegraphics[width=8cm,angle=0]{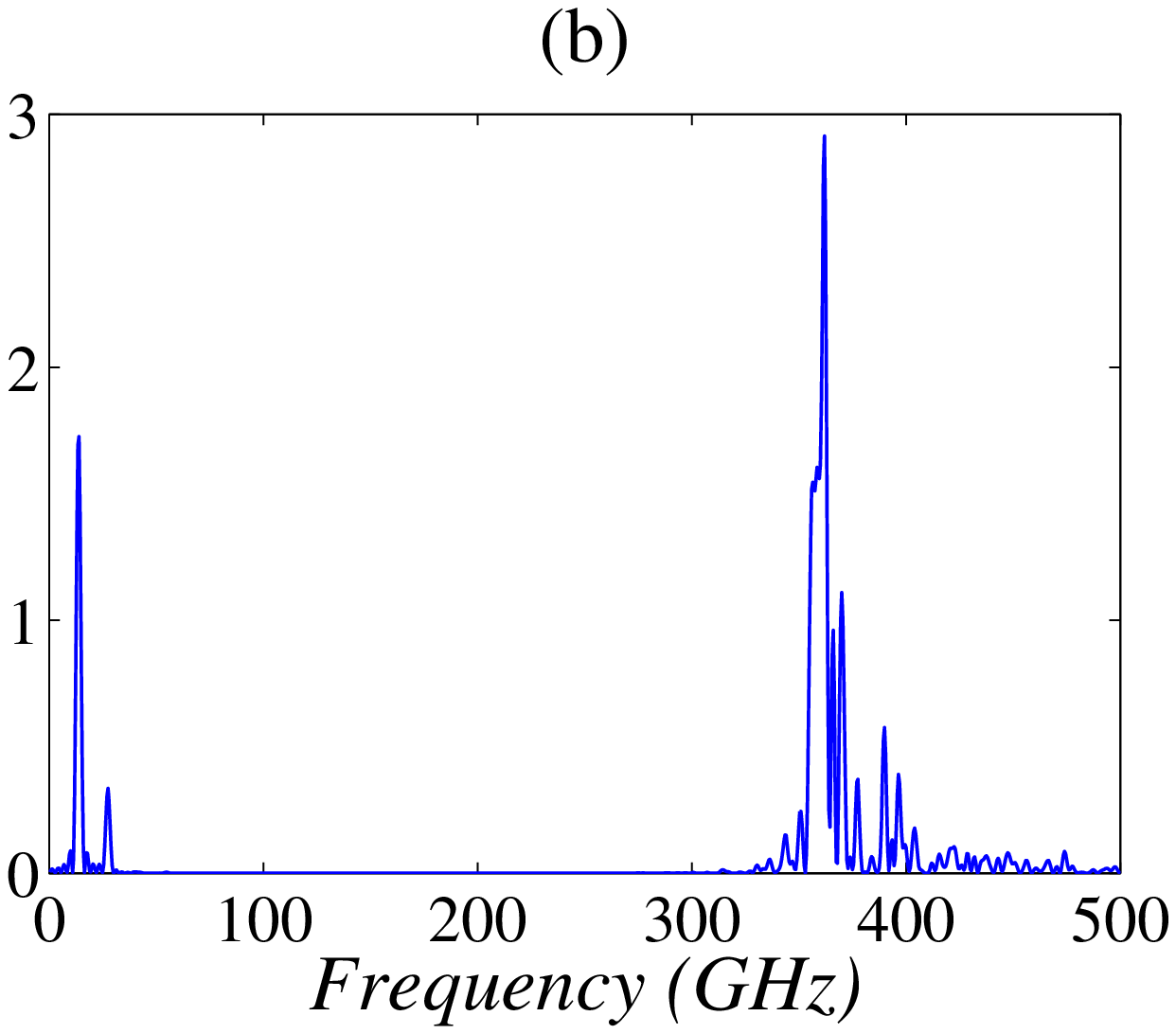}\\
\vspace{0.2cm} \caption{(a) Total current density vs time during coexisting Bloch and Gunn type oscillations at 70 K. (b) Fourier transform of the total current density showing two peaks corresponding to coexisting Bloch (0.36 THz) and Gunn type (13.8 GHz) oscillations. The zero-frequency constant corresponding to the time average of the total current density has been subtracted.} \label{fig2}
\end{center}
\end{figure}

\begin{figure}
\begin{center}
\includegraphics[width=8cm,angle=0]{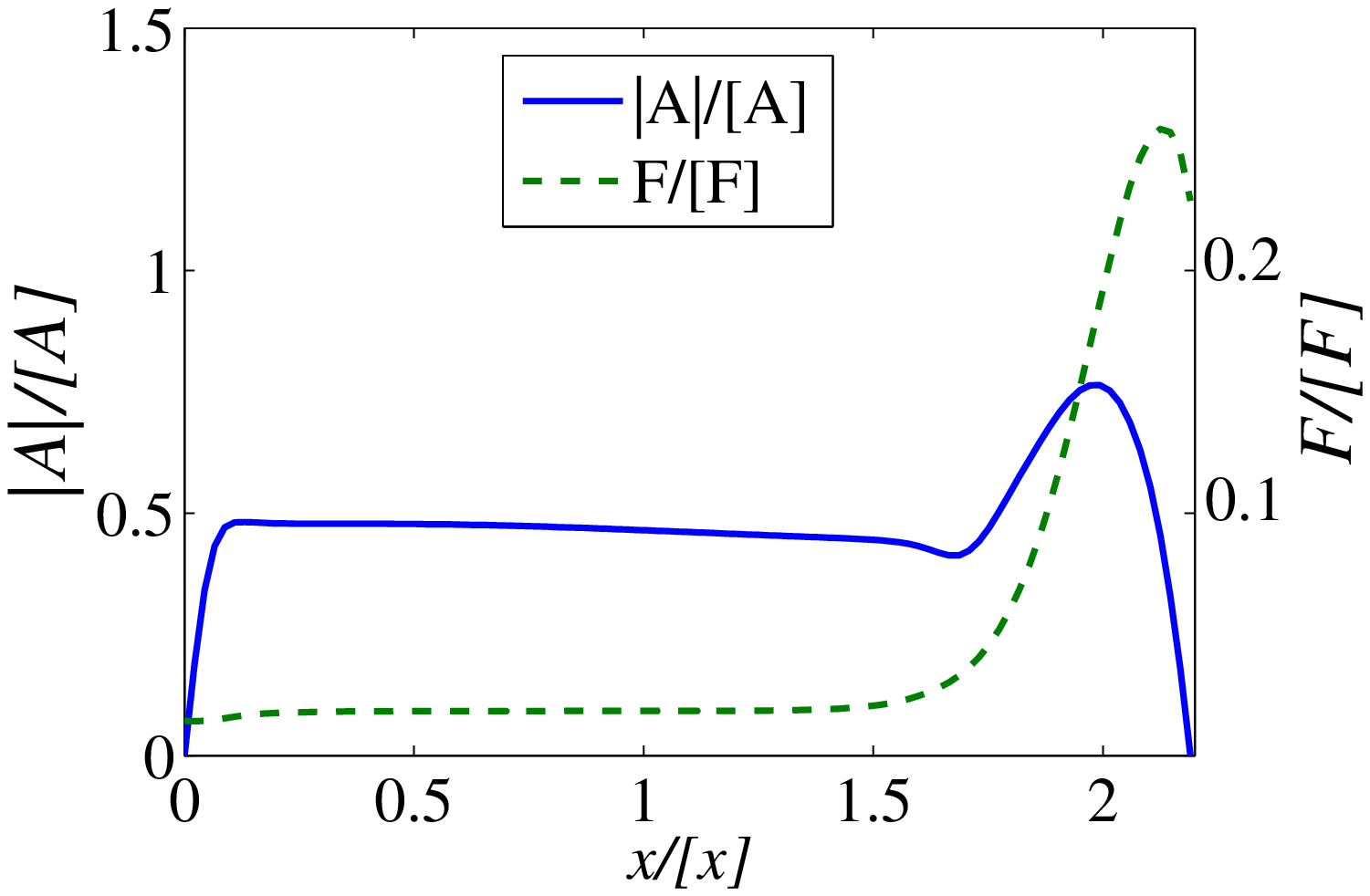}\\
\vspace{0.2cm} \caption{Modulus of the BO complex amplitude and field profiles  vs space for the stationary state at 300K.  } 
\label{fig3}
\end{center}
\end{figure}

\begin{figure}
\begin{center}
\includegraphics[width=8cm,angle=0]{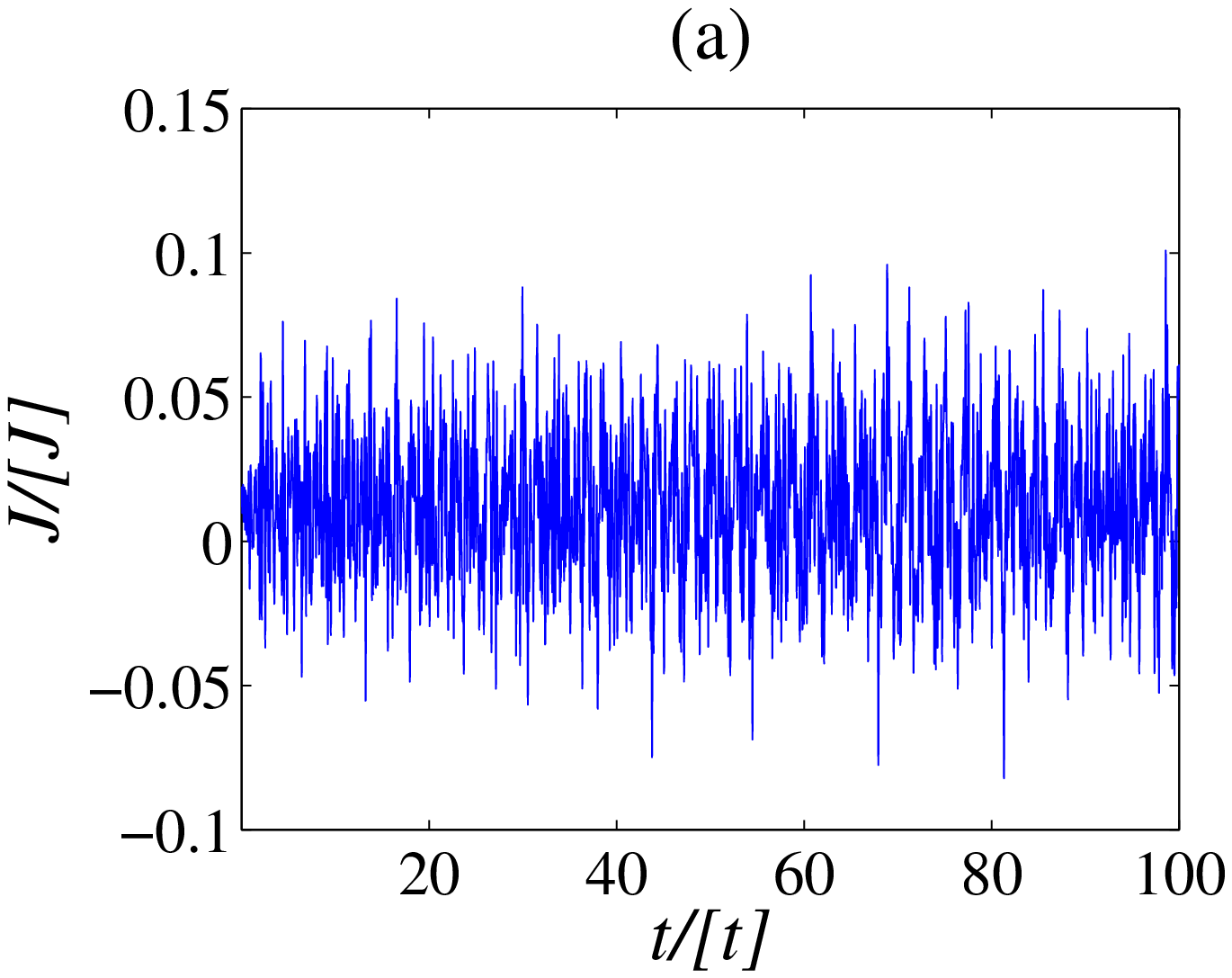}\\
\includegraphics[width=8cm,angle=0]{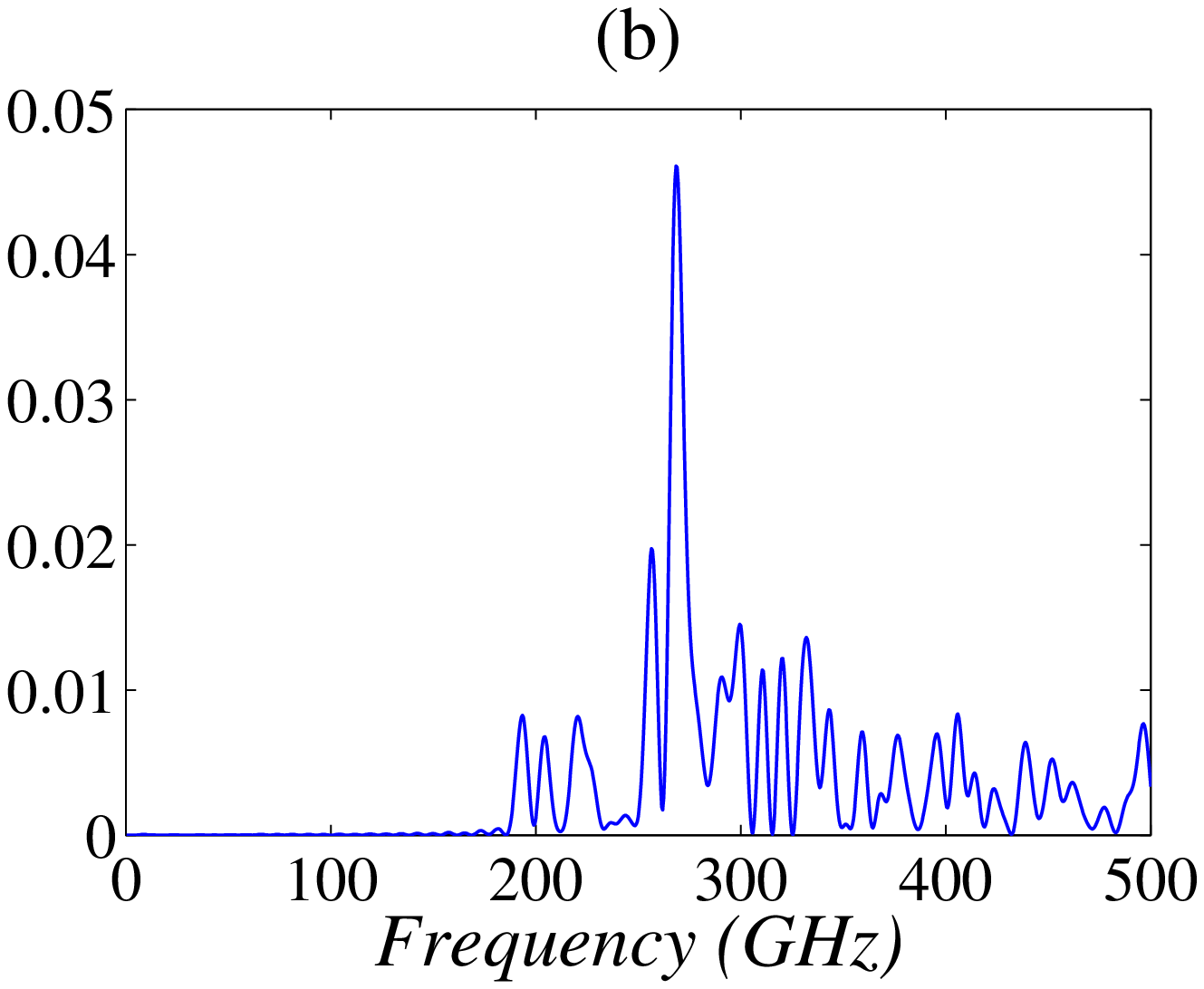}\\
\vspace{0.2cm} \caption{(a) Total current density vs time during Bloch oscillations at 300K. (b) Fourier transform of the total current density showing only one peak corresponding to  BOs (0.27 THz). The zero-frequency constant corresponding to the time average of the total current density has been subtracted.} \label{fig4}
\end{center}
\end{figure}

Figure \ref{fig3} shows the profiles of $F$ and $A$ and Fig. \ref{fig4} depicts the total current density at temperature 300K for the same values of $\alpha_{e,j}$ and the other parameters. We find BOs but not the slower Gunn type oscillations. Whether Bloch and Gunn type oscillations coexist depends on the relative size of the diffusion and convection terms in (\ref{eq129}) and (\ref{eq130}) which, in turn, are controlled by the lattice temperature. If diffusion terms are sufficiently small compared to convective terms in (\ref{eq129}) and (\ref{eq130}) (which happens for small enough lattice temperature), Gunn type oscillations mediated by EFDs in the $F$ and $A$ profiles are possible. For larger temperatures, Bloch and Gunn type oscillations cannot occur simultaneously. This latter fact was previously revealed by solving numerically a simpler version of the hydrodynamic equations with $r^{(1)}_2=0$ in (\ref{eq130}) and quite large voltage bias \cite{BAC11}. Note that the largest peak in the current spectrum occurs at a lower frequency (260 GHz) than in the case of lower lattice temperature of Fig. \ref{fig2}(b).

\section{Concluding remarks}
\label{sec:7}
We have proposed a Boltzmann-BGK kinetic equation for electron transport in miniband semiconductor superlattices. Its local equilibrium depends on electron density, mean energy and current density and therefore it oscillates periodically in time with the Bloch frequency when the mean energy and the current density do the same. This model is richer than the usual BGK models traditionally used in this field and its corresponding hydrodynamic equations may exhibit Bloch oscillations which are absent in the hydrodynamic regime of the KSS and related models. We have introduced novel singular perturbation methods to derive hydrodynamic equations describing Bloch oscillations in the limit in which collision and Bloch frequencies dominate all other terms in the kinetic equation and the collisions are almost elastic. By numerically solving the hydrodynamic equations with appropriate initial and boundary conditions, we find that nonlinearities may stabilize Bloch oscillations if the restitution coefficients are small enough. There are different scenarios depending on the lattice temperature. For sufficiently low temperature, Bloch and Gunn type oscillations mediated by electric field, current and mean energy domains may exist simultaneously for appropriate voltage ranges. These oscillations are spatially inhomogeneous and have field profiles with EFDs typical of Gunn oscillations. For larger lattice temperatures, Bloch and Gunn type oscillations do not coexist: the profiles of the electric field and the amplitude of the Bloch oscillations are independent of time but inhomogeneous in space\cite{BAC11}. As the collisions become more inelastic, the parameter range for which BOs appear shrinks and these oscillations disappear for the standard superlattices used in experiments \cite{sch98}. In the absence of BOs, the hydrodynamic equations become the known drift-diffusion system valid for inelastic collisions that may exhibit Gunn-type self-sustained oscillations due to periodic recycling of charge dipole domains for appropriate parameter values \cite{BGr05}.

\acknowledgments
LLB thanks Javier Brey for fruitful discussions about dissipative BGK models and for pointing out Ref. \cite{BMD} to him. This work has been supported by the MICINN grant FIS2008-04921-C02-01.

\setcounter{equation}{0}
\renewcommand{\theequation}{A.\arabic{equation}}
\appendix
\section{Local equilibrium distributions that cannot sustain Bloch oscillations}
\label{app0}
Let us consider the local equilibrium distribution (\ref{eq83}) which becomes 
\begin{equation}
 f^{1D\alpha}(k;n,E,J_n) = \frac{m^{*}\Delta}{2\pi\tilde{\beta}\hbar^2N_D}\, \ln\left(1+ e^{\tilde{\mu}-\tilde{\beta}+\tilde{\beta}\cos (k-k_\alpha)}\right)\!,  \label{ap01}
 \end{equation}
written in nondimensional units. Inserting this equation in (\ref{eq87}), we find
\begin{equation}
n =  f^{1D\alpha}_0=\frac{m^{*}\Delta}{2\pi\tilde{\beta}\hbar^2N_D}\, \int_{-\pi}^\pi\ln\left(1+ e^{\tilde{\mu}-\tilde{\beta}+\tilde{\beta}\cos k}\right) dk,  \label{ap02}
 \end{equation}
 after shifting the integration variable $k\to (k-k_\alpha)$. Similarly, we find
 \begin{eqnarray}
 f^{1D\alpha}_j &=& \frac{m^{*}\Delta}{2\pi\tilde{\beta}\hbar^2N_D}\, e^{-ijk_\alpha}\int_{-\pi}^\pi e^{-ijk}\ln\!\left(1+ e^{\tilde{\mu}-\tilde{\beta}+\tilde{\beta}\cos k}\right) dk\nonumber\\
 &=& \frac{m^{*}\Delta}{\pi\tilde{\beta}\hbar^2N_D}\, e^{-ijk_\alpha}\int_{0}^\pi \cos (jk)\,\ln\!\left(1+ e^{\tilde{\mu}-\tilde{\beta}+\tilde{\beta}\cos k}\right) dk.  \label{ap03}
 \end{eqnarray}
 As $\delta\to 0$, the left hand side of (\ref{ap03}) for $j=1$ becomes $Ae^{-i\theta}$ according to (\ref{eq112}), from which we obtain
 \begin{eqnarray}
|A| = \frac{m^{*}\Delta}{\pi\tilde{\beta}\hbar^2N_D}\int_{0}^\pi \ln\!\left(1+ e^{\tilde{\mu}-\tilde{\beta}+\tilde{\beta}\cos k}\right)\cos k\, dk\label{ap04}\\
 k_\alpha= \theta - \arg (A).  \label{ap05}
 \end{eqnarray}
Equations (\ref{ap03}) and (\ref{ap04}) can be solved to produce the leading order approximations of $\tilde{\mu}$ and $\tilde{\beta}$ as functions of the slowly varying quantities $n$ and $|A|$, whereas (\ref{ap05}) indicates that $k_\alpha$ varies rapidly as a shifted Bloch phase. Then (\ref{ap03}) implies that $f^{1D\alpha}_{j,l}\neq 0$ for $l=-j$ and all the other harmonics are zero. We find that $ f^{1D\alpha(0)}_{2,-1} = r^{(1)}_{2,-1} = 0$ in (\ref{eq130}) and therefore BOs are always damped for this model. 

Let us consider now (\ref{eq84}) which, for the tight binding dispersion relation, can be rewritten as 
\begin{equation}
 f^{1D\alpha}(k;n,E,J_n) = \frac{m^{*}\Delta}{2\pi\tilde{\beta}\hbar^2N_D}\, \ln\left(1+ e^{\tilde{\mu}^*-\tilde{\beta}^*+\tilde{\beta}^*\cos k-\tilde{P}_\alphaÊ\sin k)}\right)\!.  \label{ap06}
 \end{equation}
Replacing $k=k_\alpha+\xi$, we obtain 
\begin{equation}
\tilde{\beta}^*\cos k-\tilde{P}_\alphaÊ\sin k= (\tilde{\beta}^*\cos k_\alpha-\tilde{P}_\alphaÊ\sin k_\alpha)\cos\xi,  \label{ap07}
 \end{equation}
 provided 
\begin{equation}
\tilde{\beta}^*\sin k_\alpha+\tilde{P}_\alphaÊ\cos k_\alpha=0\Longrightarrow \tan k_\alpha = -\frac{\tilde{P}_\alpha}{\tilde{\beta}^*}.  \label{ap08}
 \end{equation}
Substituting (\ref{ap08}) in (\ref{ap07}), we get
\begin{equation}
\tilde{\beta}^*\cos k-\tilde{P}_\alphaÊ\sin k= \left(\tilde{\beta}^*+\frac{\tilde{P}^2_\alpha}{\tilde{\beta}^*}\right)\!Ê\cos k_\alpha)\!\cos\xi,  \label{ap09}
 \end{equation}
 which, inserted in (\ref{ap06}) yields (\ref{ap01}) with 
\begin{equation}
\tilde{\beta}=\tilde{\beta}^*+\frac{\tilde{P}^2_\alpha}{\tilde{\beta}^*}, \quad
\tilde{\mu}^*=\tilde{\mu}+\tilde{\beta}^*-\tilde{\beta}.  \label{ap10}
 \end{equation}
 This shows that (\ref{eq84}) is equivalent to (\ref{eq83}).

\setcounter{equation}{0}
\renewcommand{\theequation}{B.\arabic{equation}}
\section{Boltzmann local equilibrium distribution}
\label{app1}
In nondimensional units, the Boltzmann distribution (\ref{eq91}) satisfying $f^B_{0}=n$ is
\begin{equation}
f^B = n\,\frac{\pi\, e^{ \tilde{u}k +\tilde{\beta}\cos k}}{\int_{0}^\pi 
e^{\tilde{\beta}\cos K}\cosh(\tilde{u}K)\, dK}. \label{ap1}
\end{equation}
The first moments of this distribution can be used to calculate $\tilde{\beta}$ and $\tilde{u}$ in terms of $E$ and $J_{n}$ by solving
\begin{eqnarray}
&& \frac{\int_{0}^\pi e^{\tilde{\beta}\cos K}\cosh(\tilde{u}K)\,\cos K\, 
dK}{\int_{0}^\pi e^{\tilde{\beta}\cos K}\cosh(\tilde{u}K)\, dK}= \alpha_{e}
E_0+(1-\alpha_{e})E, \label{ap2}\\
&& \frac{\int_{0}^\pi e^{\tilde{\beta}\cos K}\sinh(\tilde{u}K)\,\sin K\, dK}{\int_{0}^\pi 
e^{\tilde{\beta}\cos K}\cosh(\tilde{u}K)\, dK} = (1-\alpha_{j})\,\frac{J_{n}}{n}.\label{ap3}
\end{eqnarray}
The left hand side of (\ref{ap3}) can be simplified by integrating the numerator by parts:
\begin{eqnarray}
\frac{\tilde{u}}{\tilde{\beta}}-\frac{e^{-\tilde{\beta} } 
\sinh(\tilde{u}\pi) }{\tilde{\beta}\,\int_{0}^\pi e^{\tilde{\beta}\cos K}
\cosh(\tilde{u}K)\, dK} = (1-\alpha_{j})\,\frac{J_{n}}{n}.\label{ap4}
\end{eqnarray}
Equations (\ref{ap2}) and (\ref{ap4}) contain the integral $\int_0^\pi e^{\tilde{\beta}\cos k}\cosh(\tilde{u}k)\, dk$ which can be calculated using the generating function \cite{AS}
\begin{eqnarray}
e^{\tilde{\beta}\cos k}= I_0(\tilde{\beta})+2\sum_{l=1}^\infty I_l(\tilde{\beta})\cos(lk),\label{ap5}
\end{eqnarray}
with the result
\begin{eqnarray}
\int_0^\pi e^{\tilde{\beta}\cos k}\cosh(\tilde{u}k)\, dk= \left[\frac{I_0(\tilde{\beta})}{\tilde{u}}+2\tilde{u}\sum_{l=1}^\infty \frac{(-1)^l}{\tilde{u}^2+l^2}I_l(\tilde{\beta})\right]\! \sinh(\tilde{u}\pi). \label{ap6}
\end{eqnarray}
From this formula we obtain 
\begin{eqnarray}
\frac{\partial}{\partial\tilde{\beta}}\ln\!\int_0^\pi e^{\tilde{\beta}\cos k}\cosh(\tilde{u}k) dk= \frac{\frac{I_1(\tilde{\beta})}{\tilde{u}}+\tilde{u}\sum_{l=1}^\infty \frac{(-1)^l}{\tilde{u}^2+l^2}[I_{l-1}(\tilde{\beta})+I_{l+1}(\tilde{\beta})]}{\frac{I_0(\tilde{\beta})}{\tilde{u}}+2\tilde{u}\sum_{l=1}^\infty \frac{(-1)^l}{\tilde{u}^2+l^2}I_l(\tilde{\beta})}\!.\,\, \label{ap7}
\end{eqnarray}
We now use (\ref{ap6}) and (\ref{ap7}) in (\ref{ap2}) and (\ref{ap4}), thereby obtaining
\begin{eqnarray}
&& \frac{I_1(\tilde{\beta})+\tilde{u}^2\sum_{l=1}^\infty \frac{(-1)^l}{\tilde{u}^2+l^2}[I_{l-1}(\tilde{\beta})+I_{l+1}(\tilde{\beta})]}{I_0(\tilde{\beta})+2\tilde{u}^2\sum_{l=1}^\infty \frac{(-1)^l}{\tilde{u}^2+l^2}I_l(\tilde{\beta})}= \alpha_{e}
E_0+(1-\alpha_{e})E, \label{ap8}\\
&& \frac{\tilde{u}}{\tilde{\beta}}\left[1 - \frac{e^{-\tilde{\beta}}}{I_0(\tilde{\beta})+2\tilde{u}^2\sum_{l=1}^\infty \frac{(-1)^l}{\tilde{u}^2+l^2}I_l(\tilde{\beta})}\right] = (1-\alpha_{j})\,\frac{J_{n}}{n}.\label{ap9}
\end{eqnarray}

\setcounter{equation}{0}
\renewcommand{\theequation}{C.\arabic{equation}}
\section{Inelastic collisions and the hyperbolic limit}
\label{sec:5}
Here we shall use the CEM to obtain equations for the electric field and the electron density in the case of inelastic collisions with $0< \alpha_{e,j}\leq 1$. In the method of multiple scales, we expand the distribution function and all its moments in powers of $\delta$ and consider slow and fast time scales. The condition that the terms in the distribution function be periodic (or, more generally, bounded as the fast time tends to infinity) in the fast time determines the modulation equations in the slow time scale. In the inelastic case, the damping coefficient $(\alpha_e+\alpha_j)/2$ in the equation for the BO amplitude is of order one. Thus the distribution function relaxes exponentially fast to a quasi-stationary function whose current and energy densities are given (to leading order) by (\ref{eq41}). This distribution is the starting point of the CEM which, in the inelastic case, is similar to that described in \cite{BEP03} and \cite{BT10}.

The leading order expression for the distribution function depends on time only through the moments $n$ and $F$ which vary on the slow time scale $t$. These moments are not expanded in powers of $\delta$. Instead, their evolution equations are expanded (as we show below), and the corresponding terms in the expansion are determined so as to keep compatibility conditions issuing from the assumptions for the distribution function. The CEM can be used to obtain reduced equations for the moments containing terms of different order in $\delta$, and this is something that the method of multiple scales cannot deliver. 

The leading-order distribution function is the solution of Eq.\ (\ref{eq82}) for 
$\delta=0$. Its Fourier coefficients are
\begin{eqnarray} 
&& \mbox{Re}f^{(0)}_{j} = \frac{\mbox{Re}f^{1D\alpha}_{j} +jF\,
\mbox{Im}f^{1D\alpha}_{j}}{1+j^2 F^2},\label{b.1}\\
&& \mbox{Im}f^{(0)}_{j} = \frac{\mbox{Im}f^{1D\alpha}_{j} - jF\,
\mbox{Re}f^{1D\alpha}_{j}}{1+j^2 F^2}.  \label{b.2}
\end{eqnarray} 
We assume that $J_{n}$ and $E$ have already acquired their quasi-stationary values after a fast decay on the time scale $\tau$. These quasi-stationary values are functions of $n$, $F$ and $\delta$ to be determined now. The Chapman-Enskog Ansatz is
\begin{eqnarray} 
&& f(x,k,t;\delta) = \sum_{m=0}^{\infty} f^{(m)}(k;F,n)\,\delta^{m} , \label{b.3}\\
&& \frac{\partial F}{\partial t}+\sum_{m=0}^\infty {\cal J}^{(m)}(F,n)\,\delta^{m}= J(t),  \label{b.4}\\
&& \frac{\partial n}{\partial t} = -\sum_{m=0}^{\infty} \frac{\partial}{
\partial x} \mathcal{J}^{(m)}(F,n)\,\delta^{m}.   \label{b.5}
\end{eqnarray} 
In (\ref{b.4}), the total current density is of course the same as its average over one period of the BOs. We have used the Poisson equation (\ref{eq86}) to obtain (\ref{b.5}). The local distribution function $f^{1D\alpha}$ is now a function of $n$ and $F$ because $J_n$ and $E$ depend now on $n$, $F$ and $\delta$. We have
\begin{eqnarray} 
f^{1D\alpha}= \sum_{m=0}^\infty f^{1D\alpha(m)}\delta^m, \label{b.6}
\end{eqnarray}
and then (\ref{eq87}) - (\ref{eq88}) yield the following compatibility
conditions:
\begin{eqnarray} 
&& f^{(0)}_{0} = f^{1D\alpha(0)}_{0} =n,   \label{b.7}\\
&&\mbox{Re}f^{(0)}_{1} = nE^{(0)},\quad \mbox{Re}f^{1D\alpha(0)}_{1} =n\, 
[\alpha_{e}E_0+(1-\alpha_{e})E^{(0)}],\label{b.8}\\
&& \mbox{Im}f^{(0)}_{1} = -J_{n}^{(0)},\quad \mbox{Im}f^{1D\alpha(0)}_{1}= - 
(1-\alpha_{j})\, J^{(0)}_{n}.  \label{b.9}
\end{eqnarray} 
Let us now find $f^{1D\alpha(m)}$ in (\ref{b.6}). Inserting (\ref{b.8}) and (\ref{b.9}) in (\ref{b.1}) and (\ref{b.2}), we obtain a system of two algebraic equations for the unknowns $nE^{(0)}$ and $J_{n}^{(0)}$ whose solution is
\begin{eqnarray} 
&& E^{(0)} = \frac{\alpha_{e}\alpha_{j} E_{0}}{\alpha_{j}\alpha_{e}+F^2},  
\label{b.10}\\
&& J_{n}^{(0)} = \frac{\alpha_{e} E_{0}n F}{\alpha_{j}\alpha_{e}+F^2}, 
\label{b.11}\\
&& f_{1}^{(0)}=nE^{(0)}-iJ_{n}^{(0)}= n\,\frac{\alpha_{e}E_{0}(\alpha_{j}-iF)}{
\alpha_{j}\alpha_{e}+F^2}.  \label{b.12}
\end{eqnarray} 
The approximate electron current density (\ref{b.11}) provides an approximate electron drift velocity vs.\ field, $v_{d}(F)=J_n^{(0)}/n$, whose maximum value is reached at
\begin{eqnarray}
v_{\rm max}=\frac{E_{0}}{2}\,\sqrt{\frac{\alpha_{e}}{\alpha_j}}=
\frac{I_{1}(\tilde{\beta}_{0})}{2I_{0}(\tilde{\beta}_{0})}\,
\sqrt{\frac{\alpha_{e}}{\alpha_{j}}},\quad 
F_{\rm max}= \sqrt{\alpha_{e}\alpha_{j}},  \label{b.13}
\end{eqnarray}
in which we have used (\ref{eq95}) to relate $E_{0}$ to the lattice temperature $1/\tilde{\beta}_{0}=2k_{B}T_{0}/\Delta$ for a Boltzmann local equilibrium. For $\alpha_{e}=\alpha_{j}=1$, (\ref{b.10}) - (\ref{b.11}) become the well-known values (\ref{eq42}) corresponding to the simple KSS-Poisson problem (\ref{eq2}) - (\ref{eq8}) with Boltzmann local equilibrium \cite{ISh87} provided $\tau_e=\sqrt{\alpha_j/\alpha_e}$. It is interesting to note that we have derived (\ref{b.10}) and (\ref{b.11}) for an unspecified general local equilibrium $f^{1D\alpha}$, not just for the Boltzmann distribution. This means that this expression for the electron drift velocity is also valid at low temperatures, when the Fermi-Dirac distribution (\ref{eq89}) is a better description, and it justifies a posteriori the use of (\ref{b.11}) to fit experimental results \cite{sch98}.
\bigskip

\noindent {\em Remark C1.} To leading order, $E$ and $J_{n}$ in the right hand sides of (\ref{ap2}) and (\ref{ap4}) can be eliminated by using (\ref{b.10}) and (\ref{b.11}), thereby  obtaining 
\begin{eqnarray}
&& \frac{\int_{0}^\pi e^{\tilde{\beta}\cos K}\cosh(\tilde{u}K)\,\cos K\, 
dK}{\int_{0}^\pi e^{\tilde{\beta}\cos K}\cosh(\tilde{u}K)\, dK} =
\alpha_{e}E_{0}\frac{\alpha_{j}+F^2}{
\alpha_{j}\alpha_{e}+F^2}, \label{b.14}\\
&&\frac{\tilde{u}}{\tilde{\beta}}-\frac{e^{-\tilde{\beta} } 
\sinh(\tilde{u}\pi) }{\tilde{\beta}\,\int_{0}^\pi e^{\tilde{\beta}\cos K}
\cosh(\tilde{u}K)\, dK}= \frac{\alpha_{e}(1-\alpha_{j}) E_{0}F}{
\alpha_{j}\alpha_{e}+F^2}.   \label{b.15}
\end{eqnarray}
Solving these two equations yield the functions $\tilde{\beta}(F)$ and $\tilde{u}(F)$. In the case $\alpha_{j}=1$, (\ref{b.15}) yields $\tilde{u}=0$ and (\ref{b.14}) becomes 
\begin{eqnarray} 
\frac{I_{1}(\tilde{\beta})}{I_{0}(\tilde{\beta})} = \frac{\alpha_{e}
(1+F^2) E_{0}}{\alpha_{e}+F^2}.   \label{b.16}
\end{eqnarray}  
\bigskip

Equations (\ref{eq87})-(\ref{eq88}) yield
\begin{eqnarray}
&&f^{(0)}_{0}=n=f^{1D\alpha(0)}_{0},\quad f^{(m)}_{0}=0,\,\mbox{for $m=1,2,
\ldots$},  \label{b.17}\\
&& \mbox{Re}f^{(m)}_{1} = nE^{(m)}, \quad \mbox{Im}f^{(m)}_{1} =- J_{n}^{(m)}. 
\label{b.18}
\end{eqnarray}
The equations for $f^{(1)}$ and $f^{(2)}$ are
\begin{eqnarray} 
\mathbb{L} f^{(1)}-f^{1D\alpha(1)} &=&  \left. - \left({\partial  f^{(0)}\over \partial t}
\right|_{0} + \sin k\,\frac{\partial f^{(0)}}{\partial x}\right),  
\label{b.19}\\
 \mathbb{L} f^{(2)}-f^{1D\alpha(2)} &=&  \left. -\left(\frac{\partial f^{(1)}}{\partial t}
 \right|_{0} + \sin k\,\frac{\partial f^{(1)}}{\partial x}\right) - 
 \left. \frac{\partial f^{(0)}}{\partial t}\right|_{1} , \label{b.20}
\end{eqnarray} 
and so on. The subscript $m=0,1$ in the right hand side of these equations means that $\partial F/\partial t$ and $\partial n/\partial t$ are replaced by $(J\delta_{m0}-\mathcal{J}
^{(m)})$ and $-\partial\mathcal{J}^{(m)}/\partial x$, respectively. In these equations,
the operator is defined by
\begin{eqnarray} 
\mathbb{L}u(k)\equiv F\, {\partial u\over \partial k}(k) + u(k) .\label{b.21}
\end{eqnarray} 
The compatibility conditions (\ref{b.17})-(\ref{b.18}) imply the following solvability conditions for the hierarchy (\ref{b.19}) and (\ref{b.20}):
\begin{equation}
(\mathbb{L} f^{(m)})_{j}= 0,\quad j=0,1.  \label{b.22}
\end{equation}

Using the solvability conditions (\ref{b.22}) for the linear hierarchy of equations, we can show that the reduced balance equations for $n$ and $F$ are obtained by inserting (\ref{b.3}) in $J_{n}=-$Im $f_1$:
\begin{equation}
J_{n}=-\sum_{m=0}^\infty \delta^m\,\mbox{Im} f^{(m)}_{1},
\quad \mathcal{J}^{(m)}= -\mbox{Im} f^{(m)}_{1}. \label{b.23}
\end{equation}
We have already calculated $\mathcal{J}^{(0)}=J_{n}^{(0)}$ to be given by Eq.\
(\ref{b.11}). To get a diffusive correction to this electron current density, we 
need to calculate Im$f^{(1)}_{1}$. From (\ref{b.19}), (\ref{b.8}), (\ref{b.9}) and (\ref{b.18}), we obtain
\begin{eqnarray} 
&&\mbox{Re}f^{(1)}_{1} = \frac{\alpha_{j}\mbox{Re} r_{1}+ F\,
\mbox{Im} r_{1}}{\alpha_{e}\alpha_{j}+F^2},\label{b.24}\\
&& \mbox{Im}f^{(1)}_{1} = \frac{\alpha_{e}\mbox{Im}r_{1} - F\,
\mbox{Re}r_{1}}{\alpha_{e}\alpha_{j}+F^2}, \label{b.25}
\end{eqnarray}  
in which 
\begin{eqnarray} 
r = \left. - \frac{\partial f^{(0)}}{\partial t}\right|_{0} - \sin k\, 
\frac{\partial f^{(0)}}{\partial x}. \label{b.26}
\end{eqnarray} 
Thus we need to find
\begin{eqnarray} 
r_{1} = - \left. \frac{\partial}{\partial x}\,\frac{n-f^{(0)}_{2}}{ 2i} - 
{\partial f^{(0)}_{1}\over\partial t}\right|_{0},  \label{b.27}
\end{eqnarray} 
in order to calculate (\ref{b.24}), i.e.,
\begin{eqnarray} 
\mathcal{J}^{(1)}= \frac{\alpha_{e}\left[\mbox{Im}\left.\frac{\partial 
f^{(0)}_{1}}{\partial t}\right|_{0} -\frac{\partial}{\partial x}\frac{n-
\mbox{Re}f_{2}^{(0)}}{2}\right] - F\left[\mbox{Re}\left. \frac{\partial 
f^{(0)}_{1}}{\partial t}\right|_{0} - \frac{1}{2}\frac{\partial}{\partial x}
\mbox{Im}f_{2}^{(0)}\right]}{\alpha_{e}\alpha_{j}+F^2}.
\label{b.28}
\end{eqnarray}  
Equation (\ref{b.12}) yields
\begin{eqnarray}
\left.\frac{\partial f^{(0)}_{1}}{\partial t}\right|_{0}&=&  \frac{-\alpha_{e}
E_{0}}{\alpha_{e}\alpha_{j}+F^2}\left[(\alpha_{j}
-iF)\frac{\partial J^{(0)}_{n}}{\partial x}\right.\label{b.29}\\
&+&\left.  n (J-J_{n}^{(0)}) \frac{2\alpha_{j}F+ i(\alpha_{e}\alpha_{j}-F^2)}{\alpha_{e}\alpha_{j}+F^2}
\right].   \nonumber
\end{eqnarray}  
The calculation of $f_{2}^{(0)}$ involves that of $f_{2}^{1D\alpha(0)}$. Using $\cos 2k=1-2\sin^2k$, $\sin 2k=2\sin k\cos k$, integrating by parts and using (\ref{ap2}), (\ref{ap4}) from \ref{app1}, and (\ref{b.14}) and (\ref{b.15}), we get 
\begin{eqnarray}
&& \frac{n-\mbox{Re}\, f^B_{2}}{2}= \frac{\alpha_{e}nE_{0}}{\tilde{
\beta}}\,\frac{1-(1-\alpha_{j})(1-\tilde{u}F)+F^2}{
\alpha_{j}\alpha_{e}+F^2},  \label{b.30}\\
&&\frac{1}{2}\,\mbox{Im}\, f^B_{2}=-\frac{n\tilde{u}}{\tilde{\beta}} -
\frac{\alpha_{e} n E_{0}}{\tilde{\beta}}\,\frac{ (1+F^2)
\tilde{u}-(1-\alpha_{j})[\tilde{u}+(1+\tilde{\beta})F]}{\alpha_{j}
\alpha_{e}+F^2}.  \label{b.31}
\end{eqnarray}
For $1-\alpha_{j}=\tilde{u}=0$, we get Im$f^B_{2}=0$ and
\begin{eqnarray}
\frac{n-\mbox{Re}\, f^B_{2}}{2}= \frac{\alpha_{e}nE_{0}}{\tilde{
\beta}}\,\frac{1+F^2}{\alpha_{e}+F^2}.  \label{b.32}
\end{eqnarray}
In this case, we obtain
\begin{eqnarray} 
&&\frac{n-\mbox{Re}f^{(0)}_{2}}{2} = \frac{n}{1+4F^2}\left[2F^2 + 
\frac{\alpha_{e} E_{0}(1+F^2)}{\tilde{\beta}\,(
\alpha_{e}+F^2)}\right],\label{b.33}\\
&& -\frac{1}{2}\mbox{Im}f^{(0)}_{2} = \frac{nF}{1+4F^2}\left[1-\frac{2\alpha_{e} E_{0}(1+F^2)}{\tilde{\beta}\,(\alpha_{e}+F^2)}\right],   \label{b.34}
\end{eqnarray}  
where $\tilde{\beta}$ is a function of $F$ found by solving the equation (\ref{b.14}).

Recapitulating, we have obtained the drift-diffusion equation (\ref{b.4}) (Amp\`ere's law) for $F$ in which $\mathcal{J}^{(0)}=J_{n}^{(0)}$ is given by (\ref{b.11}) and $\mathcal{J}^{(1)}$ is given by (\ref{b.28}) - (\ref{b.29}) and, in the particular case of a Boltzmann local equilibrium with $\alpha_{j}=1$, by (\ref{b.32}) - (\ref{b.34}). We have
\begin{eqnarray}
\frac{\partial F}{\partial t} + \frac{1}{\alpha_{j}\alpha_{e}+F^2}
\left\{ \alpha_{e}\left[E_{0}n F - \frac{\delta}{2}\frac{\partial}{\partial x}
(n-\mbox{Re}\, f^{(0)}_{2})\right.\right.\nonumber\\
+ \left.\left. \delta\,
\mbox{Im}\, \frac{\partial f^{(0)}_{1}}{\partial t}\right|_{0}\right]
+\delta F\left[ \frac{1}{2}\frac{\partial}{\partial x}\mbox{Im}\,f_{2}^{(0)} - 
\mbox{Re}\,\left. \left.
\frac{\partial f^{(0)}_{1}}{\partial t}\right|_{0}\right]\right\} =J(t),   \label{b.35}
\end{eqnarray}
where
$n=1+\partial F/\partial x$ according to the Poisson equation (\ref{eq86}). Note that the drift-diffusion equation (\ref{b.35}) coincides with the drift-diffusion equation (\ref{eq129}) when we substitute $h_{S}=\partial f_{1}/\partial t|_{0}$ given by (\ref{b.29}) (in which $J_n^{(0)}=O(\delta)$ has been neglected) and $g_{S}=f^{(0)}_{2}$ in (\ref{eq129}) with $\alpha_{e,j}=\delta\gamma_{e,j}$ according to (\ref{eq90}). Eq.\ (\ref{b.35}) for almost elastic collisions becomes
\begin{eqnarray}
&&\frac{\partial F}{\partial t} +\frac{\delta}{F^{2} +\delta^2\gamma_{j}\gamma_e}
\left[\gamma_{e}E_{0}n F+\frac{F}{2}\, \frac{\partial}{\partial x}\mbox{Im}\,f_{2,S}^{(0)}+ \delta\gamma_{e}\mbox{Im}\, \left. \frac{\partial f^{(0)}_{1}}{\partial t}\right|_{0,S}\right.\nonumber\\
&&\quad \quad \quad\left. - \frac{\delta\gamma_{e}}{2}\, {\partial\over
\partial x}(n-\mbox{Re}\, f^{(0)}_{2,S}) -F\,\mbox{Re}\,\left. 
\frac{\partial f^{(0)}_{1}}{\partial t}\right|_{0,S}\right] =\langle J\rangle_\theta,   \label{b.36}
\end{eqnarray}
where (\ref{b.29}) should be inserted and $f^{(0)}_{2}$ is given by (\ref{b.1}) and (\ref{b.2}). In Equation (\ref{eq59}) for the BO amplitude $A$, $g=f_{2}^{(0)}+\delta f_{2}^{(1)}$ with $f_{1}=f_{1,S}+A e^{-i\theta}$ and $f_{1,S}$ is given by (\ref{b.10}) and (\ref{b.11}). Therefore in the case of almost elastic collisions, if the amplitude of the Bloch oscillations decays to zero, we are left with the above written drift-diffusion problem.
\bigskip

\noindent {\em Remark C2.} Note that (\ref{eq38}) with $E^{1D}=E_0$ can be rewritten as 
\begin{eqnarray} 
f_{1,S} &=& \frac{\delta\gamma_{e} nE_0(\delta \gamma_{j}-iF)}{F^2+\delta^2 \gamma_{j} \gamma_e}\nonumber\\
&+&  \frac{\delta}{F^2+\delta^2\gamma_{j}\gamma_e}\!\left[\!\frac{F+i\delta\gamma_{e}}{2}\, \frac{\partial}{\partial x}(n-\mbox{Re}\, f_{2,0})+\frac{\delta\gamma_{j}-iF}{2}\,\frac{\partial}{\partial x} \mbox{Im}\, f_{2,0} \!\right]\nonumber\\
&+&\frac{\delta}{F^2+\delta^2\gamma_{j}\gamma_e}\left[ (iF-\delta\gamma_{j})\mbox{Re} \frac{\partial f_{1,S}}{\partial t}-(F+i\delta\gamma_{e}) \mbox{Im}\frac{\partial f_{1,S}}{\partial t}\right]\!,  \label{b.37}
\end{eqnarray} 
Our result for $h_S=\partial f_{1,S}/\partial t$ means that we have approximated $f_{1,S}$ by the first term in (\ref{b.37}). The second and third terms in (\ref{b.37}) correspond to (\ref{b.27}) which enter the $O(\delta)$ corrections (\ref{b.24}) and (\ref{b.25}) to the distribution function. Thus setting $h_S=\partial f_{1,S}/\partial t$ corresponds to $h_S=(\partial f_{1}^{(0)}/\partial t)|_0$, with $f^{(0)}_1$ given by (\ref{b.12}).



\begin{thebibliography}{9}
\bibitem{footnote1}
$F^{(0)}$ could include as an additional term an arbitrary function of $\theta$ and $t$ but this would correspond to the case of a rapidly varying voltage bias which we do not consider in this paper.
\bibitem{footnote3}
The same peak velocity and field, $v_M$ and $F_M$, as in Ref.\ \cite{sch98}, follow from the different choice $\alpha_e=0.09$ and $\alpha_j= 0.29$, which yields $\tau_e=\sqrt{29/9}$.
\end{thebibliography}
\end{document}